\documentclass{raa}
\usepackage{graphicx,times}
\usepackage{natbib,textcomp}

\begin{document}
\title{The Galactic halo magnetic field revisited}
\volnopage{{\bf 2010} Vol.\ {\bf X} No. {\bf XX}, 000--000}
\setcounter{page}{1}
\author{X. H.~Sun\inst{1, 2}
        \and
        W.~Reich\inst{2} 
        }

\institute{National Astronomical Observatories, Chinese Academy of Sciences, 
           Jia-20 Datun Road, Chaoyang District, Beijing 100012, China; 
           {\it xhsun@nao.cas.cn; xhsun@mpifr-bonn.mpg.de}\\
           \and
           Max-Planck-Institut f\"{u}r Radioastronomie, Auf dem H\"{u}gel 69, 
           53121 Bonn, Germany; {\it wreich@mpifr-bonn.mpg.de}\\
           \vs \no
           \small Received [year] [month] [day]; accepted [year] [month] [day] }

\abstract{Recently, \citet{srwe08} published new Galactic 3D-models of magnetic 
fields in the disk and halo of the Milky Way and the distribution of cosmic-ray 
electron density by taking into account the thermal electron density model 
NE2001 by \citet{cl02,cl03}. The models successfully reproduce observed 
continuum and polarization all-sky maps and the distribution of rotation 
measures of extragalactic sources across the sky. However, the model parameters 
obtained for the Galactic halo, although reproducing the observations, seem 
physically unreasonable: the magnetic field needs to be significantly stronger 
in the Galactic halo than in the plane and the cosmic-ray distribution must be 
truncated at about 1~kpc to avoid excessive synchrotron emission from the halo. 
The reason for these unrealistic parameters was the low scale-height of the 
warm thermal gas of about 1~kpc adapted in the NE2001 model. However, this 
scale-height seemed well settled by numerous investigations. Recently, the 
scale-height of the warm gas in the Galaxy was revised by \citet{gmcm08} to 
about 1.8~kpc, by showing that the 1~kpc scale-height results from a systematic 
bias in the analysis of pulsar data. This implies a higher thermal electron 
density in the Galactic halo, which in turn reduces the halo magnetic field 
strength to account for the observed rotation measures of extragalactic 
sources. We slightly modified the NE2001 model for the new scale-height and 
revised the \citet{srwe08} model parameters accordingly: the strength of the 
regular halo magnetic field is now 2~$\mu$G or lower, and the physically 
unrealistic cutoff in $z$ for the cosmic-ray electron density is removed. The 
simulations based on the revised 3D-models reproduce all-sky observations as 
before.
}

\titlerunning{Galactic halo magnetic field revisited}
\authorrunning{X. H. Sun \& W. Reich}

\maketitle

\section{Introduction}

It is important to obtain a realistic model of the Galactic magnetic field, the 
cosmic-ray electron and the thermal electron density distribution for the 
understanding of the physical processes in the magnetized interstellar medium. 
Relativistic electrons lose their energy in the Galactic magnetic field by 
polarized synchrotron emission. Synchrotron emission fluctuations are the 
primary contamination for the analysis of cosmic microwave background 
observations. The Galactic magnetic field causes deflections of ultrahigh 
energy cosmic-rays, which must be corrected for to identify their origin. The 
magnetized thermal interstellar medium generates rotation measures (RMs) in the 
direction of extragalactic sources, which must be properly subtracted. However, 
it is very challenging to obtain realistic Galactic 3D-models because of our 
position inside of the Galactic disk and the limited number of all-sky surveys 
needed for this task.

The results from earlier modelling of the Galactic magnetic field, relativistic 
electron and thermal electron densities have been reviewed by \citet{srwe08}. 
Most of the models were derived from selected data sets only and do not agree 
with other observations. \citet{srwe08} made the first attempt to establish 
3D-emission models (SRWE08 models hereafter) aiming to properly represent all 
relevant radio observations available such as RMs of extragalactic sources, and 
total intensity and polarization all-sky surveys. \citet{jfwe09} and 
\citet{jlb+10} followed that attempt and additionally invoked a quantitative 
comparison with the observations. However, \citet{jfwe09} did not take into 
account total intensity maps and \citet{jlb+10} so far presented a 2D disk 
model. \citet{srwe08} relied on a qualitative comparison between simulations 
and observations, which seems at the present stage sufficient to constrain 3D 
models describing the global Galactic properties. A quantitative comparison 
needs to separate local large-scale features and more distant anomalies from 
the all-sky maps, which is a very ambitious task by its own and beyond our 
scope.

The SRWE08 models are based on the thermal electron density model NE2001 by 
\citet{cl02,cl03}, which uses a scale-height of about 1~kpc. To account for the 
RMs of extragalactic sources at high latitudes a strong regular halo magnetic 
field of up to about 10~$\mu$G was required. Subsequently, the cosmic-ray 
electron density distribution needs to be cut in $z$\footnote{The cylindrical 
coordinate ($R$, $\phi$, $z$) is defined as: $R$ is the Galactocentric radius, 
$\phi$ is the azimuth angle starting from $l=0\degr$ and increasing in 
anticlockwise direction, and $z$ is the distance to the Galactic plane.}
at 1~kpc below and above the Galactic plane to avoid excessive polarized 
emission at high latitudes. However, both the strong halo field and the 
$z$-truncation of the cosmic-ray electron density are physically unrealistic. 
As already argued by \citet{srwe08} this problem could be solved by increasing 
the thermal electron density scale-height by a factor of about 2. Recently, a 
scale-height of about 1.8~kpc was derived by \citet{gmcm08}, which is very 
close to our prediction. This motivates us to update the SRWE08 models.  

The paper is organized as follows: we briefly describe the method and available 
observations in Sect.~2, revise the SRWE08 models in Sect.~3, and 
summarize our results in Sect.~4.

\section{The method and available observations}

We follow the same method to obtain 3D-emission models as already detailed by 
\citet{srwe08}. The thermal electron density model NE2001 was the basis for the 
SRWE08 models and needs revision to adapt for the larger scale-height by 
\citet{gmcm08}. The thermal electron density $n_e$ and the regular magnetic 
field component $B_\parallel$ along the line-of-sight $l$ were obtained from 
RMs of extragalactic sources, ${\rm RM} \sim n_{e}B_\parallel l$. The random 
field and the cosmic-ray electron density were constrained by total intensity 
and polarization all-sky maps. The Hammurabi code by \citet{wjr+09} was used to 
simulate all-sky emission, which was then compared to the observations. Despite 
recent efforts the RMs of extragalactic sources are still sparsely distributed 
and are likely influenced by local structures on large scales in an unknown way,
so that a quantitative $\chi^{2}$-fit will not determine the halo parameters 
more reliably compared to our qualitative approach.

We used RMs from the Canadian Galactic Plane Survey \citep[CGPS, ][]{btj03} and 
the Southern Galactic Plane Survey \citep[SGPS, ][]{bhg+07} to constrain the 
magnetic field in the Galactic disk. For the halo, previous RM measurements 
collected by \citet{hmb+97} and the Effelsberg L-band RM survey (Han et al. in 
prep.) were used by \citet{srwe08}. Recently, \citet{tss09} obtained RMs 
towards 37,543 extragalactic sources by re-processing the polarization data 
from the NVSS survey \citep{ccg+98}. This is the largest RM data set available 
so far. \citet{mgh+10} pointed out that individual NVSS RMs might have large  
errors, as they were derived from two frequencies only. The NVSS RMs are partly 
not reliable towards the plane because the large RMs there are usually beyond 
the ambiguity limits as explained by \citet{tss09}. However, this data set 
should describe Galactic large-scale structures well beyond the plane. 
Therefore we included these RM data to constrain the magnetic field in the halo.
 The RMs for the region of $100\degr<l<120\degr$ and $-5\degr<b<20\degr$ from 
the CGPS extension (Brown et al. in prep.) were also used. 

The 408~MHz total intensity all-sky survey \citep{hssw82} and the WMAP 
five-years 22.8~GHz polarization map \citep{hwh+09} both trace Galactic 
synchrotron emission and were used to constrain the total magnetic fields and 
cosmic-ray electron density. The WMAP five-years MEM free-free emission 
template \citep{gbh+09} served as a crosscheck for the thermal electron density 
distribution model. 

\section{Modelling revisited}

\subsection{Thermal electron density}

NE2001 is a 3D-model describing the diffuse Galactic thermal gas 
\citep{cl02,cl03}. The thermal electron density distribution consists of 
several components: a thin disk, a thick disk, spiral arms and in addition a 
large number of thermal source complexes, which is, however, not complete. The 
ionized gas at high latitudes primarily attributes to the thick disk component 
with a scale-height of about 1~kpc and a mid-plane density of about 
0.034~cm$^{-3}$. This scale-height was determined by fitting DM~$\sin|b|$ 
versus $|z|$ of pulsars, which have either measured parallaxes or are 
associated with globular clusters at known distances. ${\rm DM}\sim n_e l$ is 
the dispersion measure of a pulsar at a distance $l$. 

The NE2001 model was widely used, although shortcomings were noted soon such as 
the decrease of $z$ with pulsar distance \citep{kbm+03,lfl+06}. Recently, 
\citet{gmcm08} convincingly demonstrated that many DMs of low-latitude pulsars 
are influenced by local structures such as \ion{H}{II} regions along the 
line-of-sight and derived a scale-height of about 1.8~kpc by including only 
pulsars at $|b|\geq40\degr$. The mid-plane density derived was 0.014~cm$^{-3}$. 
\citet{gmcm08} also found that the filling factor of the thermal gas, which was 
not addressed in the NE2001 model, grows exponentially in the range of 
$0\leq|z|\leq1.4$~kpc with a scale-height of 0.7~kpc. This differs from the 
filling factors obtained by \citet{bmm06} and \citet{bm08}. 

\begin{figure}[!htbp]
\centering
\resizebox{0.45\textwidth}{!}{\includegraphics[angle=-90]{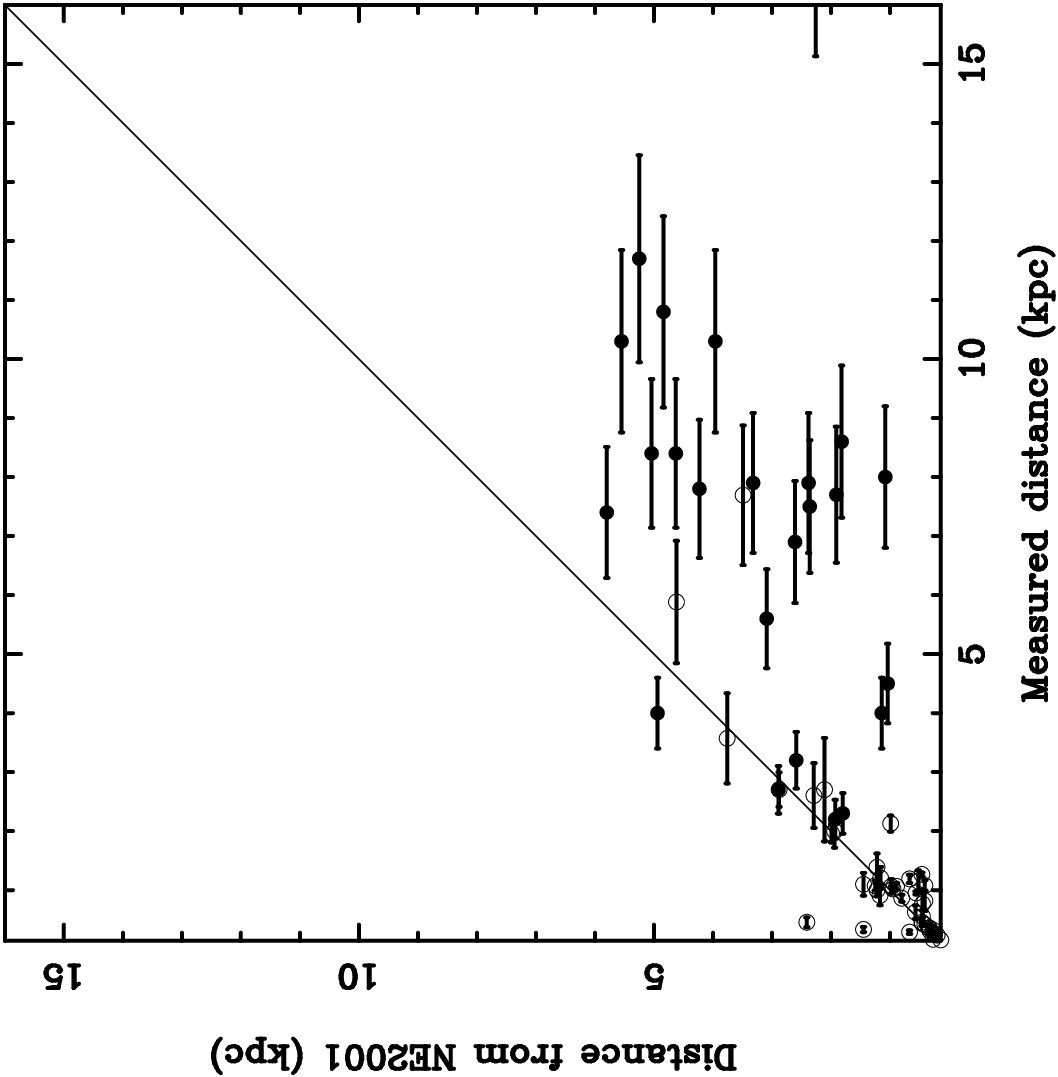}}
\resizebox{0.45\textwidth}{!}{\includegraphics[angle=-90]{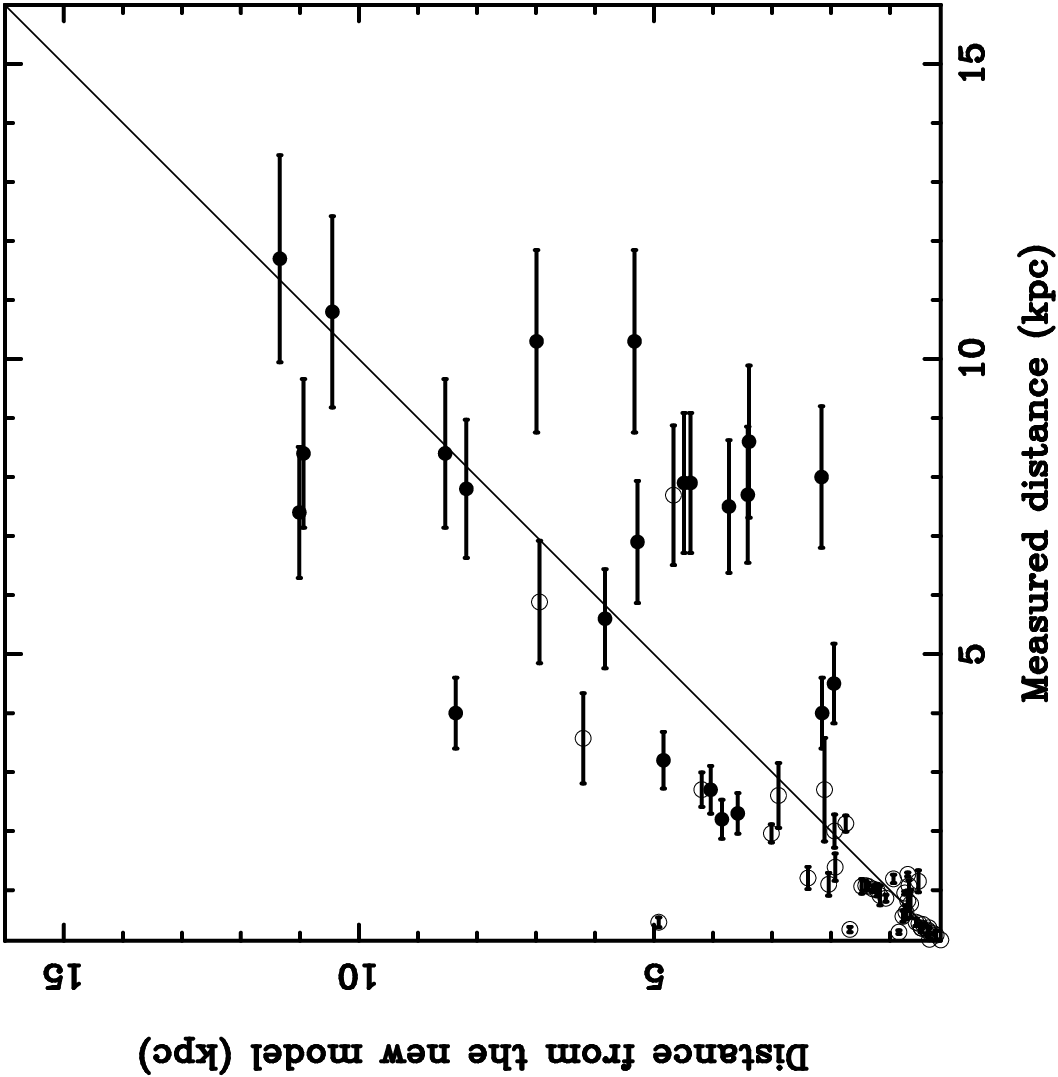}}
\caption{Measured distances versus distance estimates from the NE2001 model 
(left panel) and the modified NE2001 model (right). The measurements from 
parallaxes are indicated by open circles and those from the associated globular 
clusters by filled circles. The lines mark the case that the measured and 
estimated distances are equal.}
\label{dpsr}
\end{figure}

In the following we modify the NE2001 model by replacing the scale-height and 
the mid-plane density for the thick disk component by those provided by 
\citet{gmcm08}. We investigated the modified NE2001 model by comparing the 
distances of pulsars estimated from the model by their quoted DMs with that 
from independent measurements such as parallax and the associated globular 
clusters. The up-to-date parallax measurements for pulsars were collected by 
\citet{vlm10}. We obtained DMs of these pulsars from the ATNF pulsar 
database\footnote{http://www.atnf.csiro.au/research/pulsar/psrcat/}. Data for 
pulsars in globular clusters were taken from the 
web-page\footnote{http://www.naic.edu/{\texttildelow}pfreire/GCpsr.html} 
maintained by Paulo Freire, where the distances of the clusters and the DMs of 
pulsars are listed. The results from the original as well as from the modified 
NE2001 model are shown in Fig.~\ref{dpsr}. Although there is a considerable 
scatter, the distances of pulsars estimated from the modified NE2001 model 
agree better with the observations than the original NE2001 model, 
especially for pulsars with distances larger than about 6~kpc.

We do not claim that our modified NE2001 model is a complete substitute 
for the NE2001 model in describing the diffuse ionized gas in the Galaxy.
Other NE2001 components such as the thin disk and the arm parameters need 
to be revised correspondingly to interpret properly interstellar scattering and 
scintillation. A new model, NE2008, accounting for all new relevant 
observation is currently been developed (Jim Cordes, private communication).  

\begin{figure}[!htbp]
\centering
\resizebox{0.45\textwidth}{!}{\includegraphics[angle=-90]{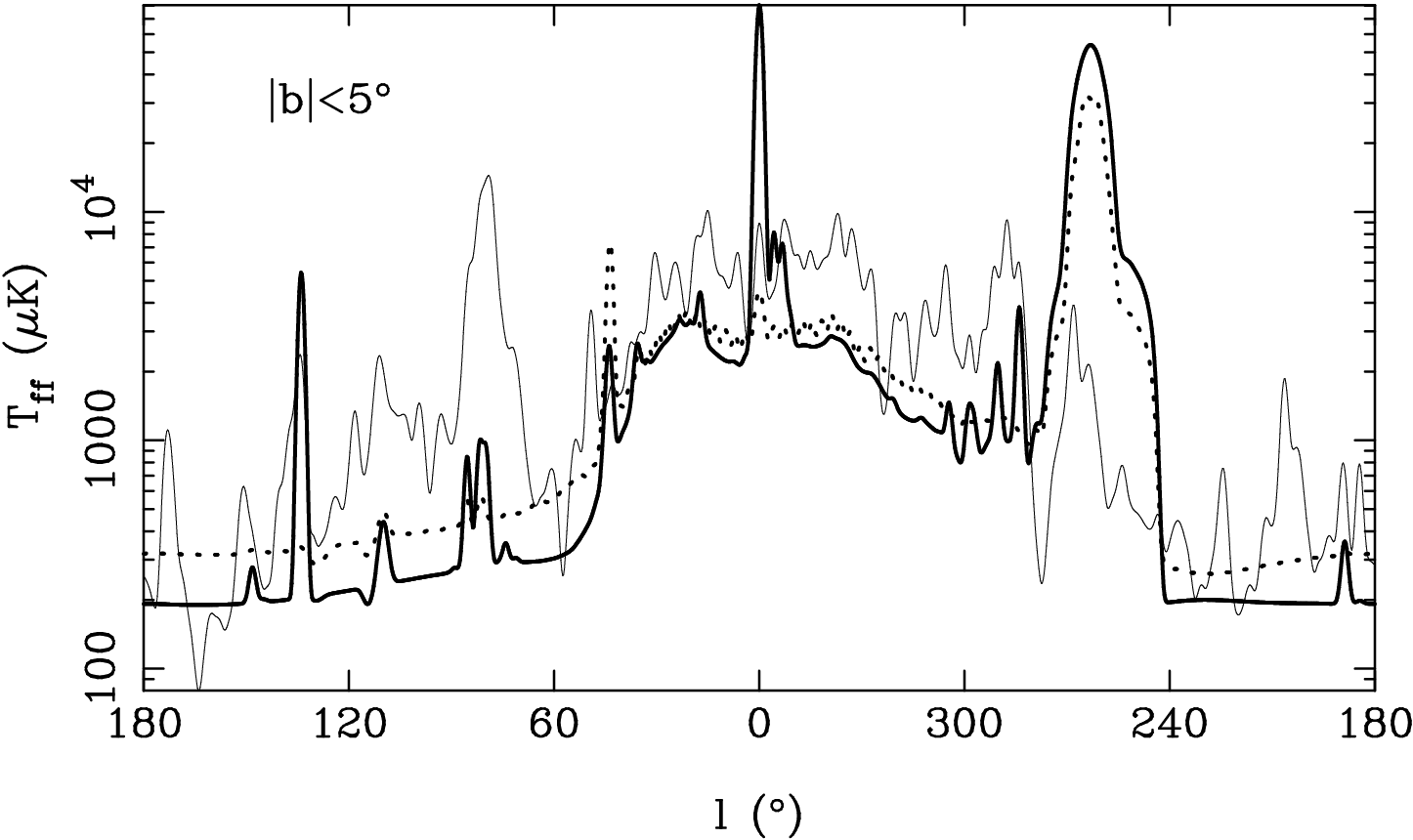}}
\resizebox{0.45\textwidth}{!}{\includegraphics[angle=-90]{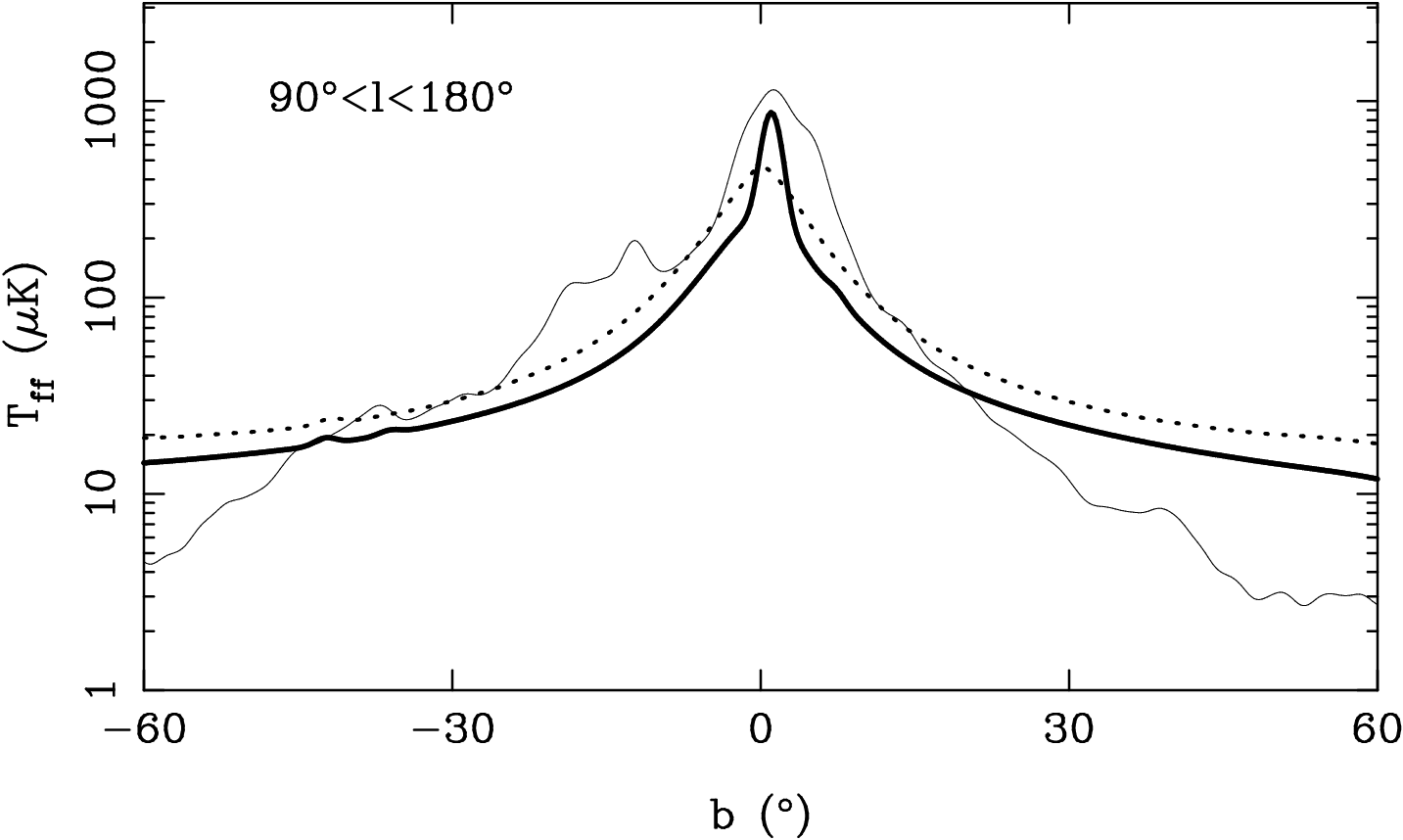}}
\resizebox{0.45\textwidth}{!}{\includegraphics[angle=-90]{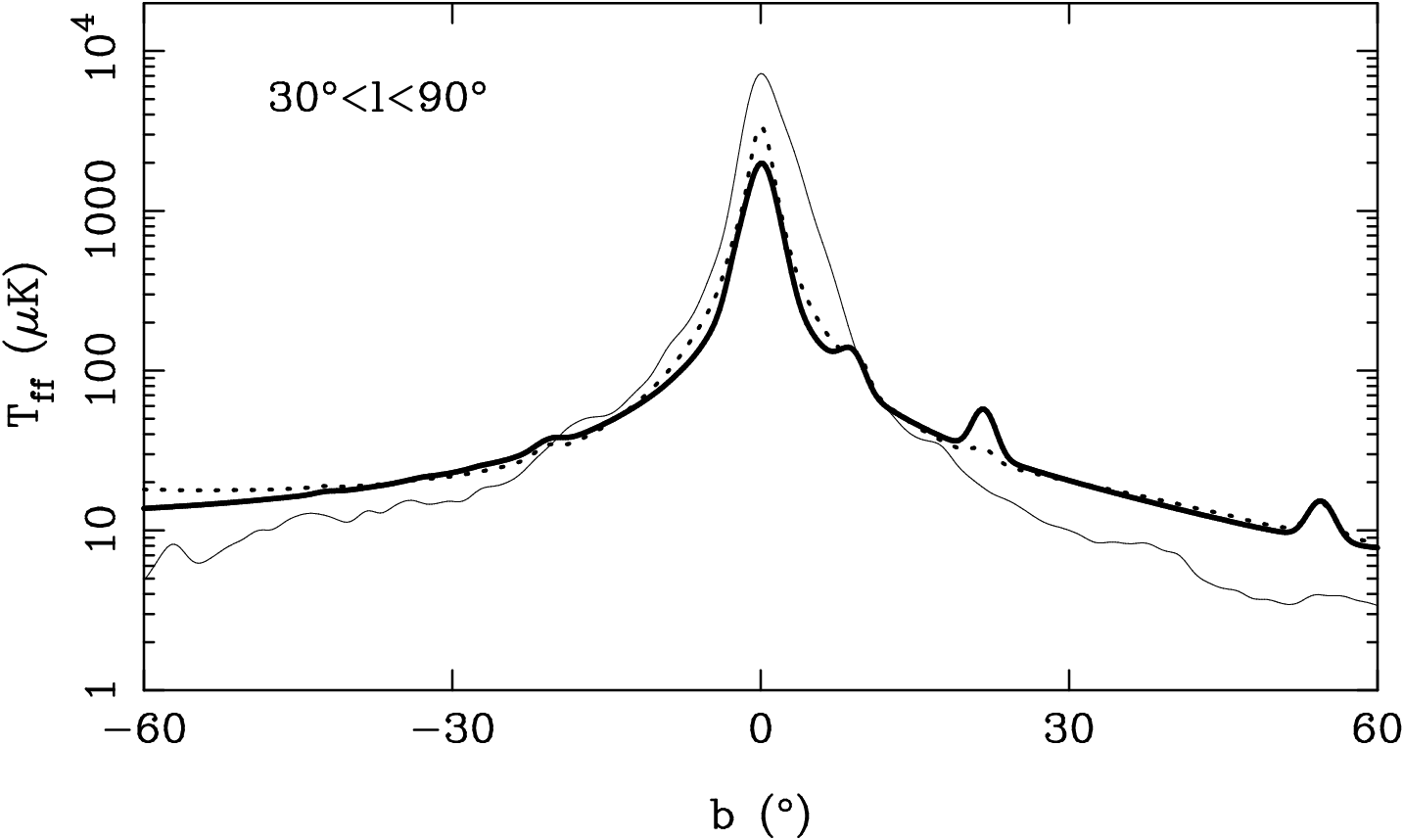}}
\resizebox{0.45\textwidth}{!}{\includegraphics[angle=-90]{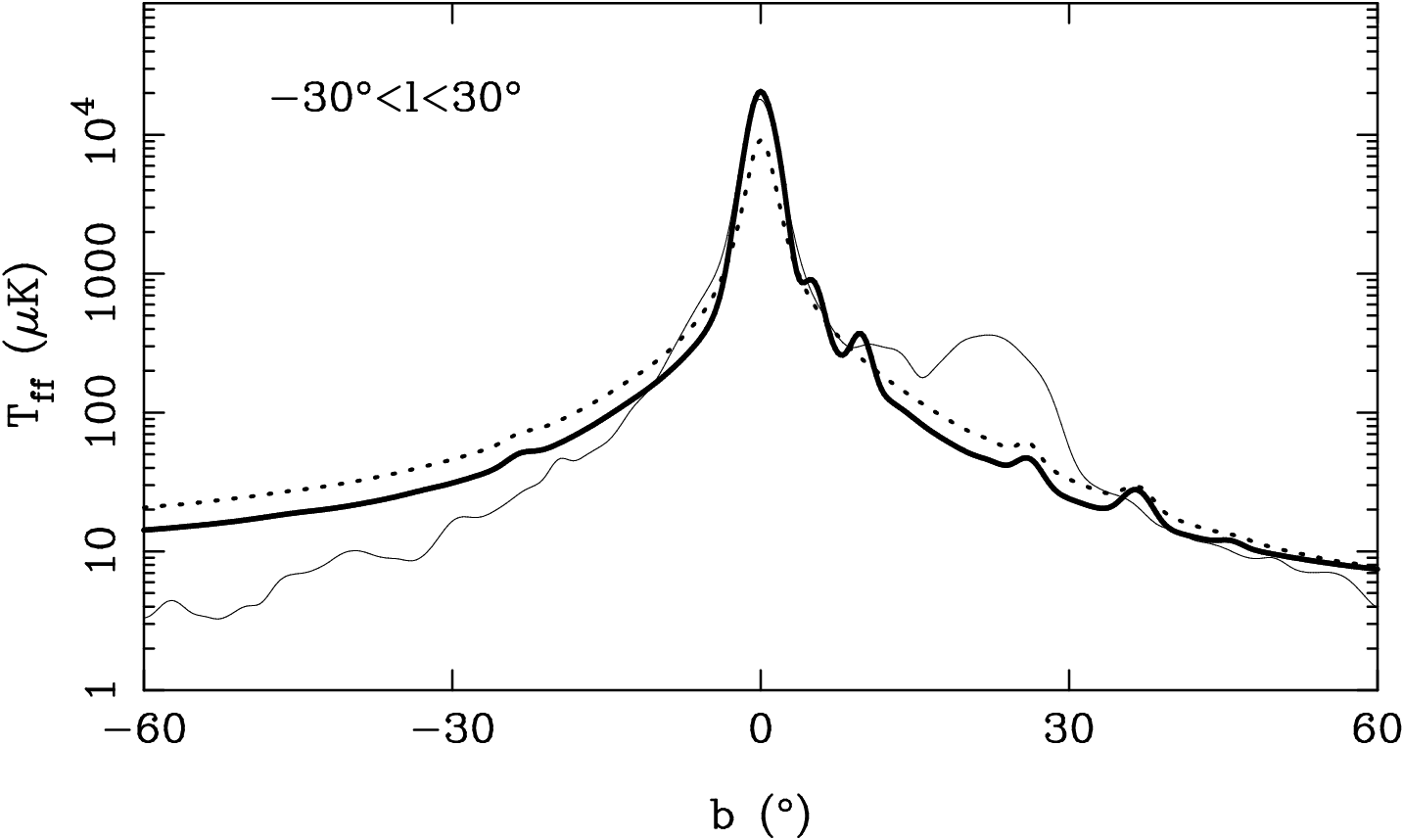}}
\caption{Longitude and latitude profiles from the simulated and the template 
free-free emission maps at 22.8~GHz. The thin solid lines are from the WMAP 
five-years MEM template, the thick solid lines from the modified NE2001 model 
with the filling factors by \citet{gmcm08}, and the dotted lines from the 
original NE2001 model with filling factors derived by \citet{bmm06}.
}
\label{ffgp}
\end{figure}

With the modified NE2001 thermal electron density model and the filling factors 
by \citet{gmcm08} as input we simulated  the all-sky free-free emission at 
22.8~GHz. This is compared to the result by \citet{srwe08} based on the 
original NE2001 model and the filling factors by \citet{bmm06}. The WMAP 
five-years MEM thermal emission template \citep{gbh+09} is shown for 
comparison. Slices extracted from the above mentioned maps are shown in 
Fig.~\ref{ffgp}. Note that the lower envelope of the observed emission is to be 
compared with the simulated profiles, which do not include all individual 
\ion{H}{II} regions nor other local features. Along the Galactic plane thermal 
emission from the modified NE2001 model is lower than that by \citet{srwe08}, 
but still agrees with the observations. The reduction of thermal emission 
results from the smaller mid-plane electron density of 0.014~cm$^{-3}$ in the 
modified model and 0.034~cm$^{-3}$ in the NE2001 model. 

The revised scale height by \citet{gmcm08} was not accepted everywhere.  
\citet{sw09} used the same data as \citet{gmcm08}, but invoked a different 
fitting scheme, where an adjustable patchiness error was included to assure that
$\chi^2_\nu\sim 1$. The fit by \citet{gmcm08} yielded $\chi^2_\nu\sim5$. 
\citet{sw09} obtained a scale-height of about 1.4~kpc instead of 
1.8~kpc by \citet{gmcm08}. Their corresponding mid-plane density is then about 
0.016~cm$^{-3}$. We run simulations by modifying the thick disk components of 
the NE2001 model with the values by \citet{sw09}, and found the results do not 
vary significantly from those with a scale-height of 1.8~kpc, which we 
presented above. New pulsar data became recently available being almost 
consistent with the 1.8~kpc scale-height also when applying the fitting method 
used by \citet{sw09} (Bryan Gaensler, private communication).

\subsection{Regular magnetic field properties}

It is a customary to split the regular Galactic magnetic field into a disk and 
a halo component. The configuration of the disk fields were usually classified 
into three types, (1) axi-symmetric spiral (ASS); (2) bi-symmetric spiral (BSS);
(3) following the spiral arms. The commonly used halo field 
patterns are toroidal, poloidal and a combination of both. The disk and halo 
fields and in particular their direction were mainly constrained by observed 
RMs. 

\subsubsection{The disk field}\label{diskb}

Early models for the disk magnetic field frequently used to study the 
propagation of ultrahigh energy cosmic-rays do not agree with the RMs observed 
from extragalactic sources \citep{srwe08} or from pulsars \citep{njkk08}.
\citet{srwe08} have proposed three new models for the disk magnetic field 
configuration: ASS+RING with reversals in the Galactocentric rings, BSS, and 
ASS+ARM with reversals in arms. All models predict RMs consistent with the 
available data. Reversals of the large-scale magnetic field have been subject 
of a controversial debate for a long time. Recently, \citet{nk10} concluded 
that the magnetic field in the fourth Galactic quadrant exhibits reversals at 
the arm-interarm interfaces, which confirmed the earlier results obtained by 
\citet{hml+06}. However, given the uncertain distances of pulsars and the 
difficulties to properly assess the influence of foreground structures such as 
\ion{H}{II} regions, supernova remnants, Faraday Screens or the giant local 
loops, this topic clearly needs further investigations. 

\begin{figure}[!htbp]
\centering
\resizebox{0.45\textwidth}{!}{\includegraphics[angle=-90]{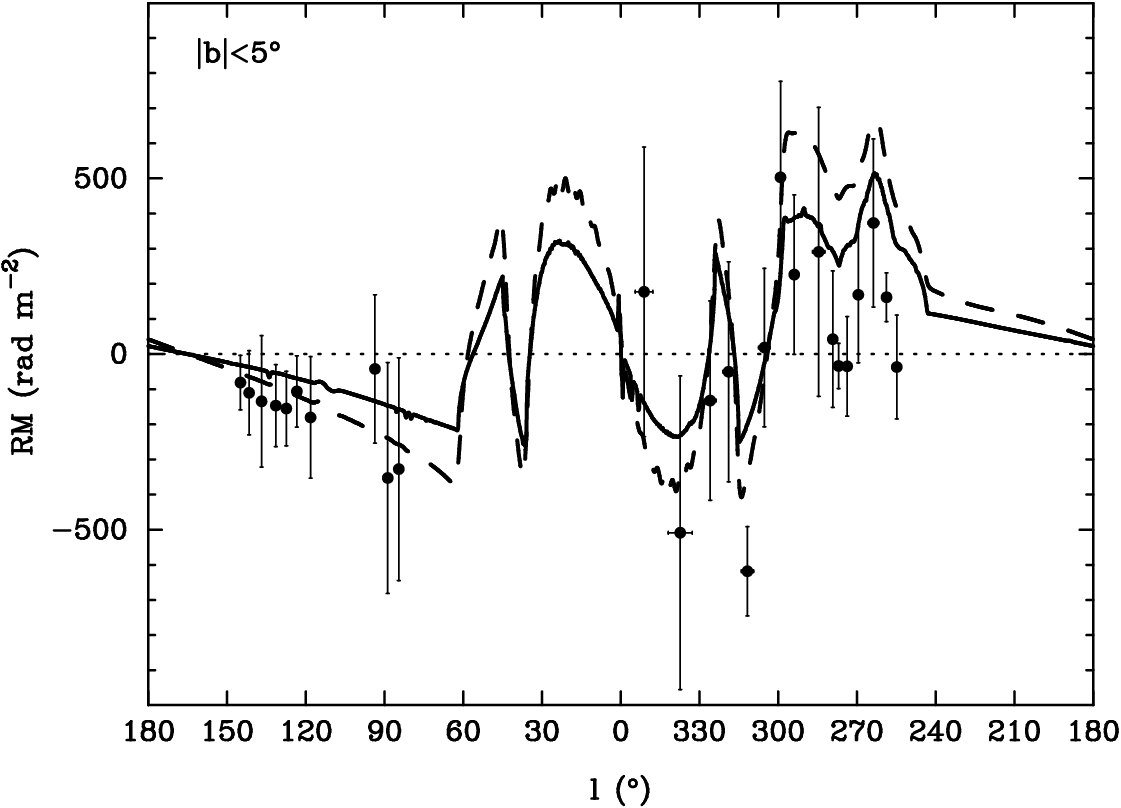}}
\resizebox{0.45\textwidth}{!}{\includegraphics[angle=-90]{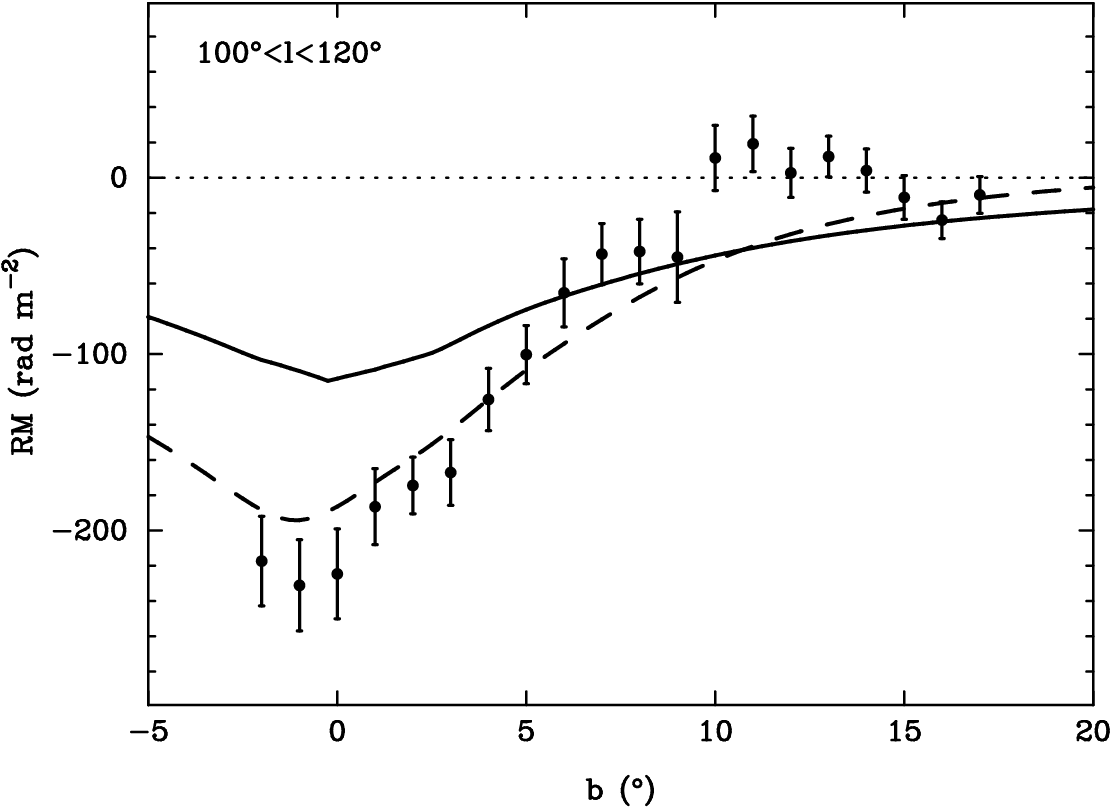}}
\resizebox{0.45\textwidth}{!}{\includegraphics[angle=-90]{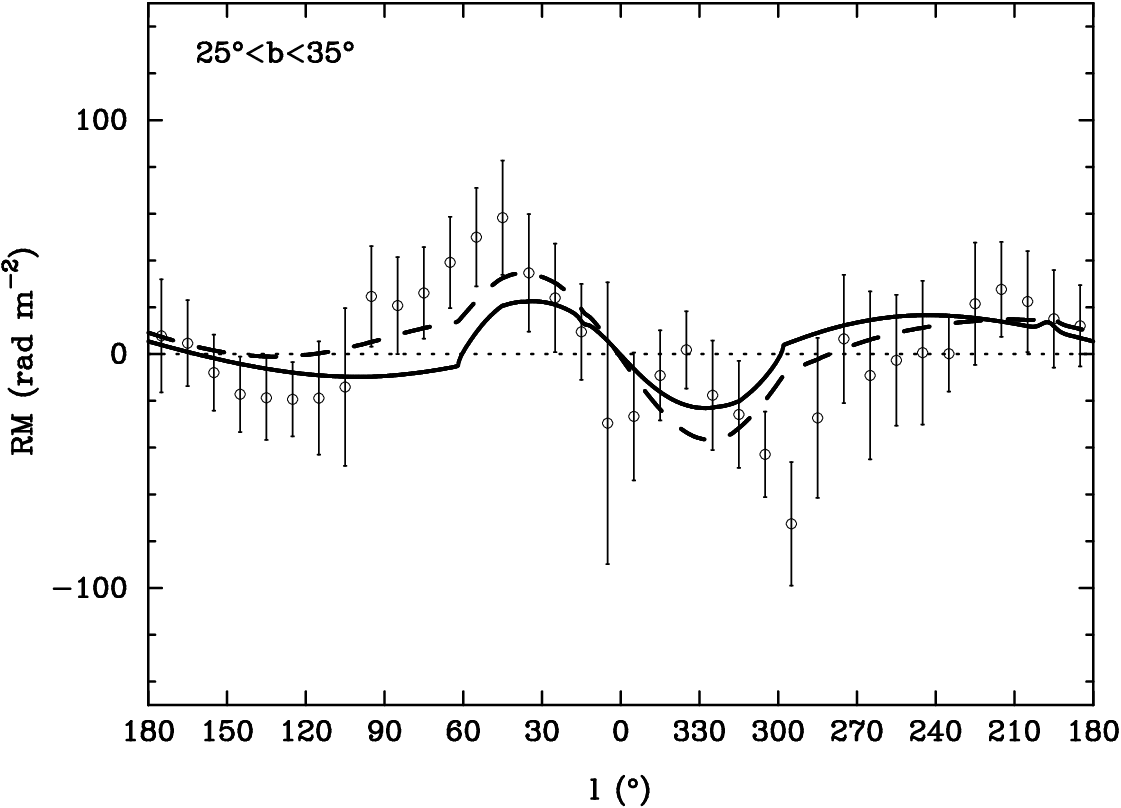}}
\resizebox{0.45\textwidth}{!}{\includegraphics[angle=-90]{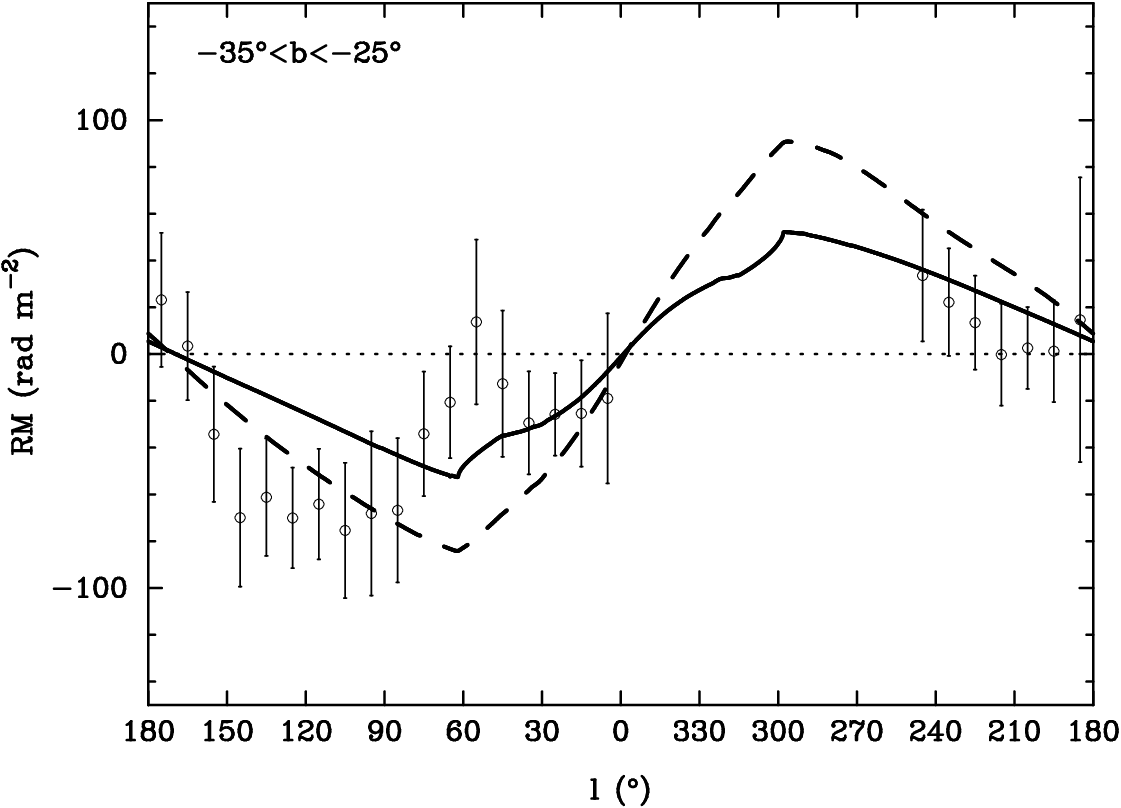}}
\resizebox{0.45\textwidth}{!}{\includegraphics[angle=-90]{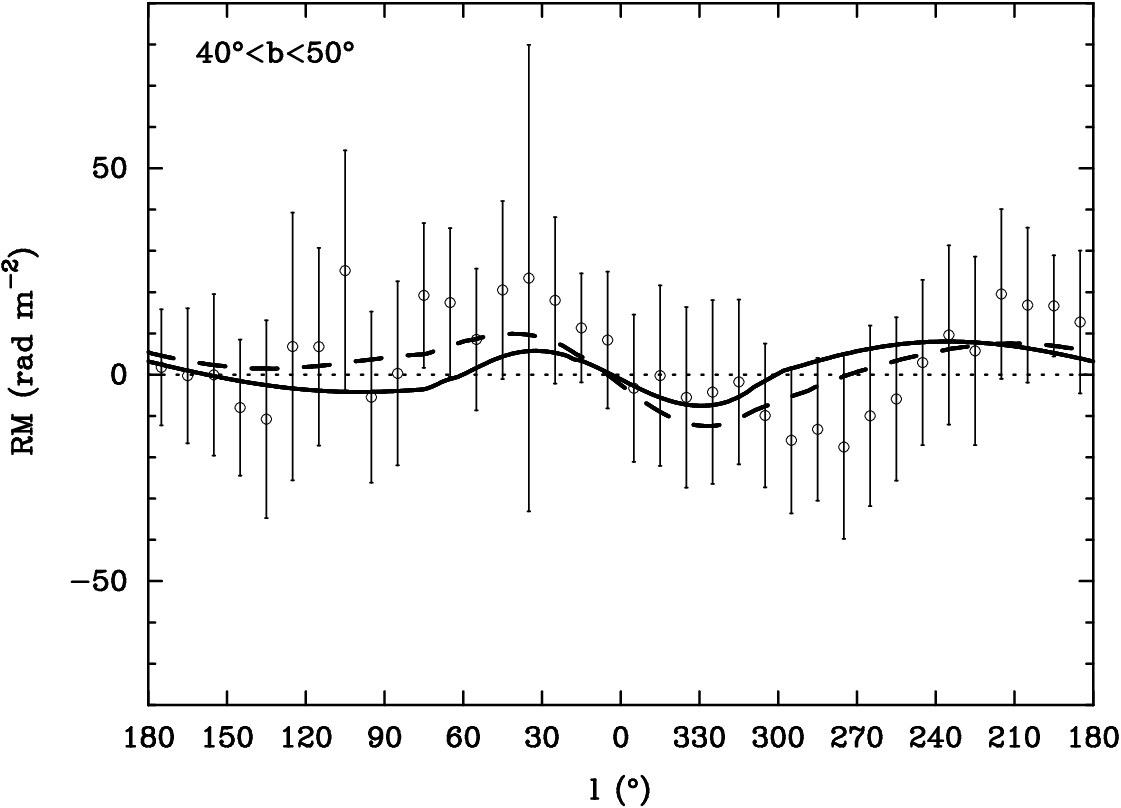}}
\resizebox{0.45\textwidth}{!}{\includegraphics[angle=-90]{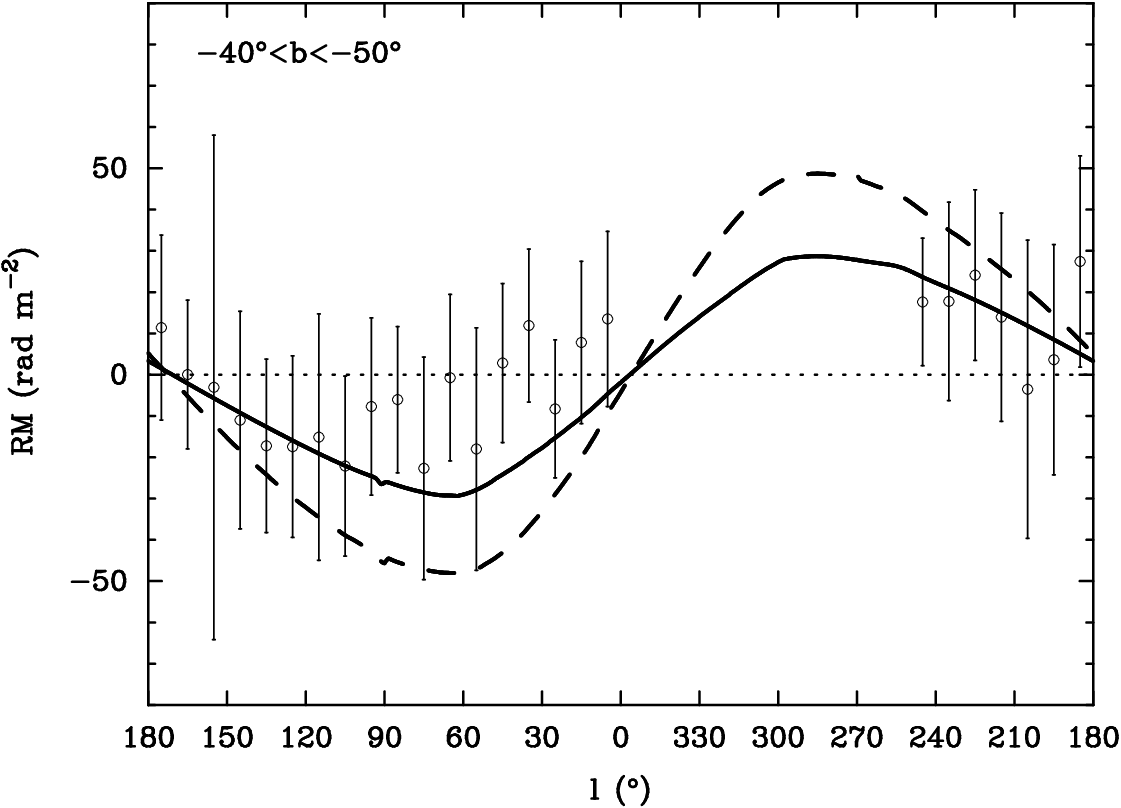}}
\caption{RM profiles along Galactic longitudes and latitudes. The solid lines 
are from the present ASS+RING model, and the dashed lines from the 
corresponding SRWE08 model. The RMs in the plane (upper-left panel) are from 
the CGPS and SGPS \citep{btj03,bhg+07}. RMs in the area $100\degr<l<120\degr$ 
and $-5\degr<b<20\degr$ (upper-right panel) are from the CGPS extension 
(Jo-Anne Brown, in prep.), and in the halo regions from NVSS sources 
\citep{tss09}.}
\label{rm}
\end{figure}

Here we kept the disk field of the SRWE08 models unchanged and simulated the RM 
map with the ASS+RING model as an example, which was recently further supported 
by new RM measurements \citep{eb09}. The RM profile along the Galactic plane 
taken from the maps is compared to the binned CGPS and SGPS RM data 
(Fig.~\ref{rm}: upper-left panel). The RM profile based on the modified NE2001 
model and the ASS+RING disk field basically reproduce all the observed features.
In general, the RM magnitudes from the present model are smaller than those 
modelled by \citet{srwe08}. For example, the RM gradient in the range of 
$80\degr\leq l\leq150\degr $ is now more shallow than before. To reach an 
agreement between the updated and the SRWE08 models we need to increase either 
the disk magnetic field strength or the thermal electron density. The 
local field strength of about 2~$\mu$G is determined by the RM/DM ratio of 
pulsars \citep[e.g. ][]{hml+06}. \citet{bssw03} pointed out that the estimate 
of magnetic field strength by the RM/DM ratio is biased if there exists a 
correlation between the thermal electron density and the magnetic field 
strength. However, \citet{wkr+09} performed simulations which showed that the 
RM/DM ratio correctly represents the mean strength of the magnetic field. 
Although there remain uncertainties on the turbulent properties of the 
interstellar medium, we consider the 2~$\mu$G local field strength as very 
robust. By increasing the mid-plane thermal electron density by about a factor 
of two it is about the same as in the NE2001 model and the RM profile from the 
present model is consistent with that from the SRWE08 model. However, the 
${\rm DM}\sin|b|$ versus $z$ could not be fitted well with the larger 
mid-plane electron density and a scale-height of 1.8~kpc. Possibly the 
exponential description of the thermal electron density is oversimplified or an 
additional electron density component exists, which contributes just near to 
the plane. We expect this point will be solved by the expected NE2008 model. 

\subsubsection{The halo field}

The asymmetric distribution of RMs in longitude and latitude relative to the 
Galactic plane and the Galactic centre indicates that the Galactic halo field 
has opposite signs below and above the plane. \citet{srwe08} have described the 
toroidal halo field $B_\phi(R,z)$ following \citet{ps03} by, 
\begin{equation}
B_\phi(R,z)={\rm sign(z)}B_0
\frac{1}{1+\left(\displaystyle{\frac{|z|-z_0}{z_1}}\right)^2}
\frac{R}{R_0}\exp\left(-\frac{R-R_0}{R_0}\right),
\end{equation}
where ${\rm sign}(z)$ takes the signs of $z$. 

To account for the revised thermal electron density model, we obtained new 
parameters for the halo field as: $z_0=1.5$~kpc, $z_1=0.2$~kpc for $|z|<z_0$ 
and otherwise $z_1=4$~kpc, $B_0=2$~$\mu$G, and $R_0=4$~kpc. From new all-sky 
simulations with added halo and disk magnetic fields, we obtained the RM 
latitude profile for the longitude range of $100\degr<l<120\degr$ 
(Fig.~\ref{rm}: the upper-right panel) to be compared with RM data from the 
CGPS high-latitude extension (Brown et al., in prep.). Near the Galactic plane 
the simulated RM profile does not agree with the observations, which again is 
caused by the too small mid-plane thermal electron density in our modified 
NE2001 model. The deviations vanish by increasing the thermal electron density 
by a factor 2 as mentioned in Sect.~\ref{diskb} to reconcile the RM difference 
in the plane. 

We also obtained average RM longitude profiles and show them for 
$25\degr<b<35\degr$ and $40\degr<b<50\degr$, as well as their southern 
counterparts in Fig.~\ref{rm} (middle and lower panels). These profiles were 
compared with the RMs from NVSS sources \citep{tss09} binned in $10\degr$ 
longitude intervals. RM profiles from new simulations generally reproduce the 
observed RM anti-symmetry with smaller amplitudes compared to the SRWE08 
models. 

There is evidence based on polarization data from the Global Magneto-Ionic 
Medium Survey (GMIMS)\footnote{https://www.astrosci.ca/users/drao/gmims}, 
currently running at the DRAO 26-m telescope at L-band \citep{wlc09,wlh10},
that magnetized filaments are associated with a very local and therefore very 
extended \ion{H}{I} shell located above the Galactic plane in the direction of 
the Galactic centre with large positive and negative RMs in the first and forth 
quadrant, respectively \citep{wfl10}. Very likely this bubble 
explains the RM deviations between the current halo model and the RM 
measurements seen in Fig.~\ref{rm} for the area $30\degr<l<60\degr$ and 
$25\degr<b<35\degr$. This finding does not question the present halo model, as 
its asymmetry above the plane of 10-20~rad~m$^{-2}$ is smaller than that 
attributed to the magnetized filaments. In the region $45\degr<l<60\degr$ and 
$-60\degr<b<-20\degr$ \citet{tss09} derived positive RMs for distinct small 
areas (see their Fig.~4), which partly cancel the in general negative RMs in 
this area and cause deviations between simulations and observations. These 
small areas with positive RMs may be associated to southern Loop~I filaments. 

It is important to keep in mind that we observe the halo magnetic field in 
superposition with the disk field. Because the toroidal magnetic field above 
the plane is directed opposite to the disk field, the intrinsic RM asymmetry 
from the halo model is reduced. Vice versa, below the plane disk and halo field 
add and the halo asymmetry becomes more evident (Fig.~\ref{rm}). It is 
therefore very important to obtain a large set of RM data for the southern sky, 
especially for the region of $240\degr<l<360\degr$, to prove the predicted halo 
RM asymmetry. We note that the northern sky area with large negative RMs within 
$60\degr<l<140\degr$ below the Galactic plane towards $b\sim-40\degr$, named 
region~A by \citet{sk80}, might be influenced by Loop~II, which is generally 
believed to be a local shell structure. 

\citet{nm10} suggested that a vertical field component needs to be added to the 
SRWE08 models to better describe deflections of ultrahigh energy cosmic-rays. 
This vertical field component has a strength of about 0.2~$\mu$G at the solar 
position consistent with that derived by \citet{hq94}. We performed simulations 
including a vertical field of that strength as proposed by \citet{ps03} and 
found the results just marginally deviate from the simulations without this 
component. This means that such a weak vertical field cannot be constrained by 
our simulations. Recently, \citet{mgh+10} obtained RMs of numerous 
extragalactic sources towards the northern and southern Galactic pole and 
convincingly demonstrated that there does not exist a coherent vertical field 
at the Sun's position. Therefore we do not include any vertical magnetic field 
component in our models.
 
\subsection{Random magnetic fields and the cosmic-ray electron density}

\subsubsection{The random magnetic field component}

\citet{srwe08} and \citet{sr09} have invoked isotropic random magnetic fields 
with a Gaussian distribution to simulate all-sky maps. Simulations of high 
angular resolution patches of the sky follow a power-law spectrum. 
\citet{jlb+10} have included an additional magnetic field component, called 
``ordered" component, which is a regular field with numerous small-scale field 
reversals. Unlike isotropic random fields, this component contributes to total 
and polarized emission, which in turn reduces the cosmic-ray electron density 
to match observations. The ``ordered" magnetic field does not increase RM, but 
increases its scatter. Although this ``ordered" component is clearly helpful to 
reach consistency between simulations and observations, the spatial scales of 
its reversals are entirely unclear and there is no proof so far on its 
existence in our Galaxy at all. We therefore do not include an ``ordered" 
component in our simulations. 

\begin{figure}[!htbp]
\centering
\resizebox{0.45\textwidth}{!}{\includegraphics[angle=-90]{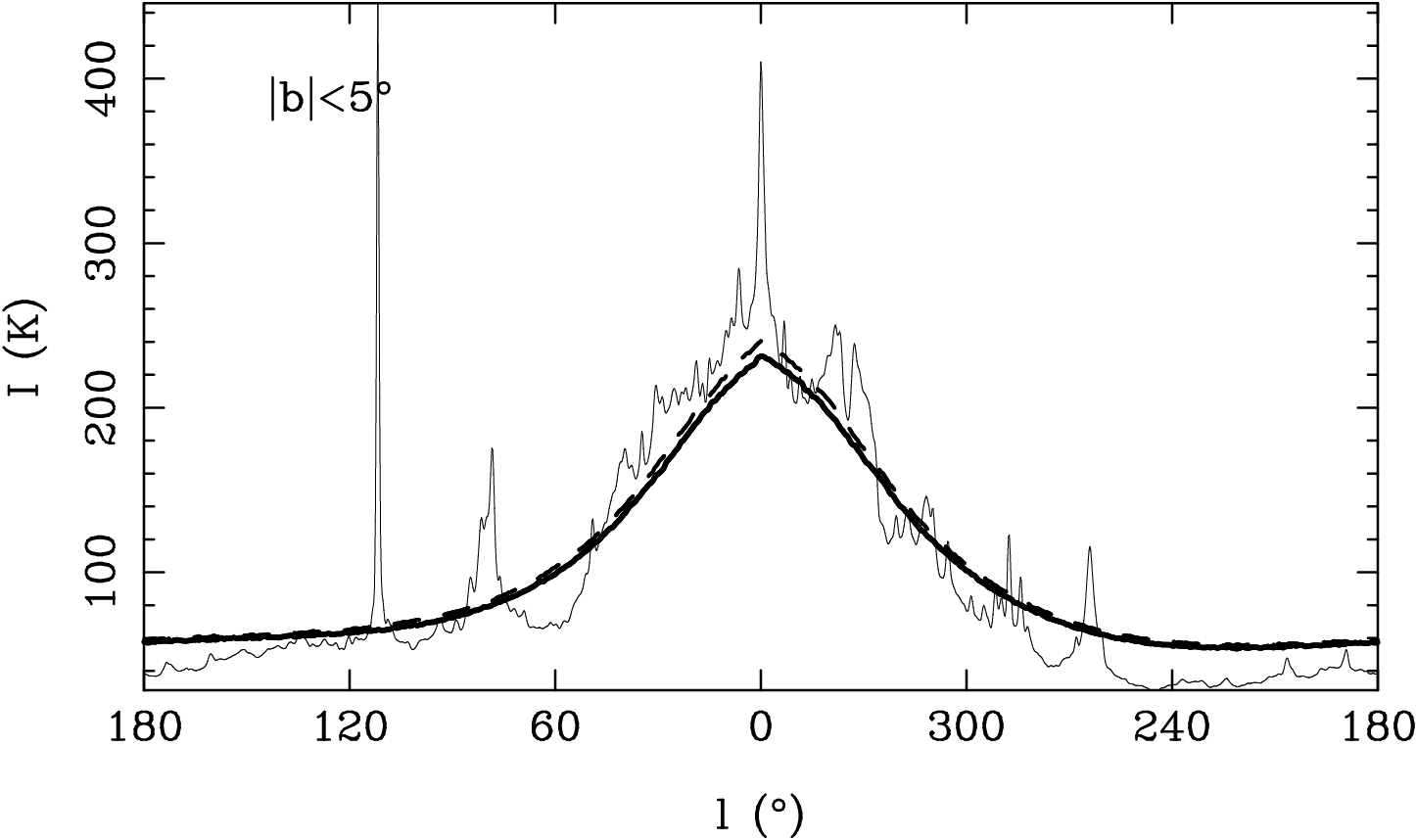}}
\resizebox{0.45\textwidth}{!}{\includegraphics[angle=-90]{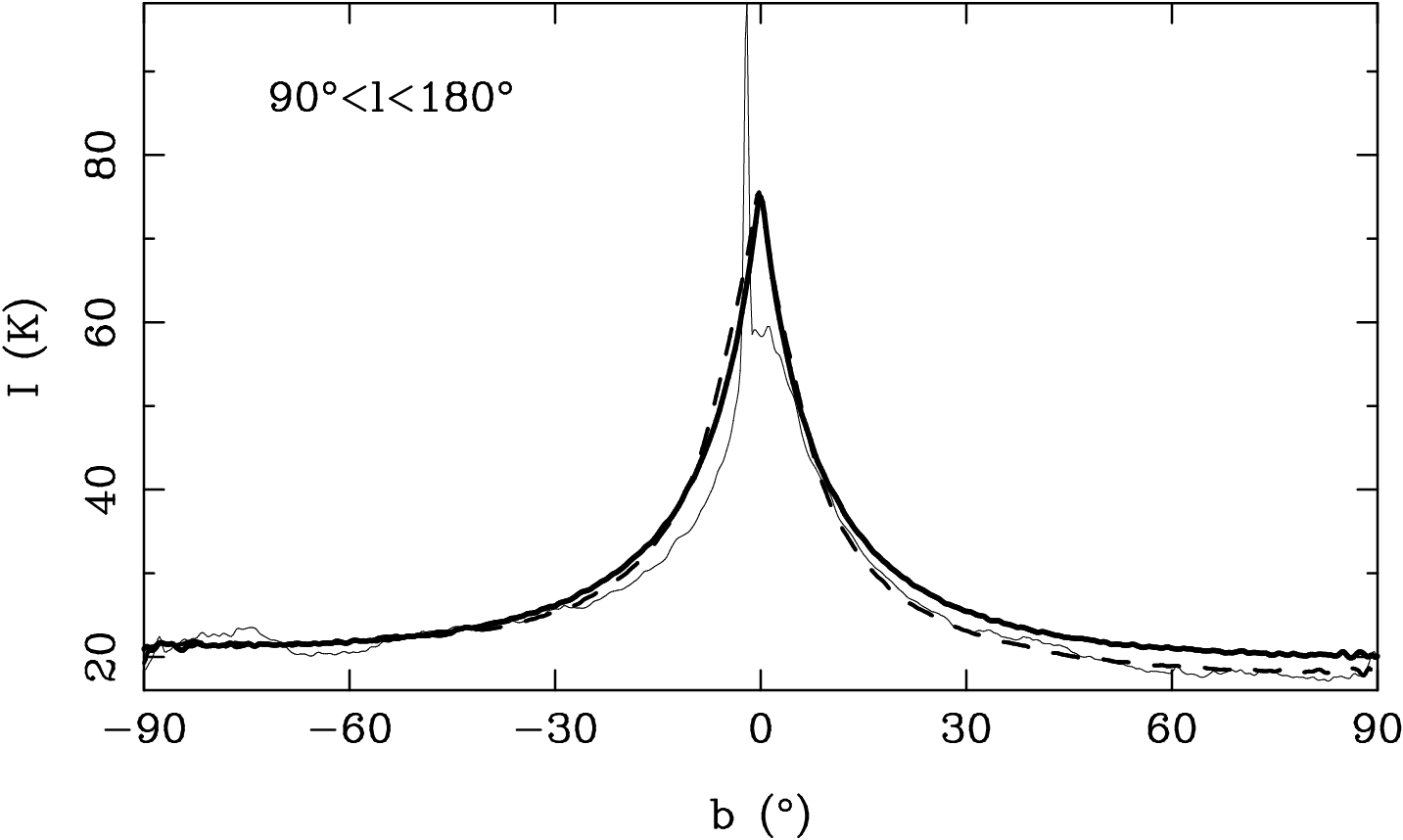}}
\resizebox{0.45\textwidth}{!}{\includegraphics[angle=-90]{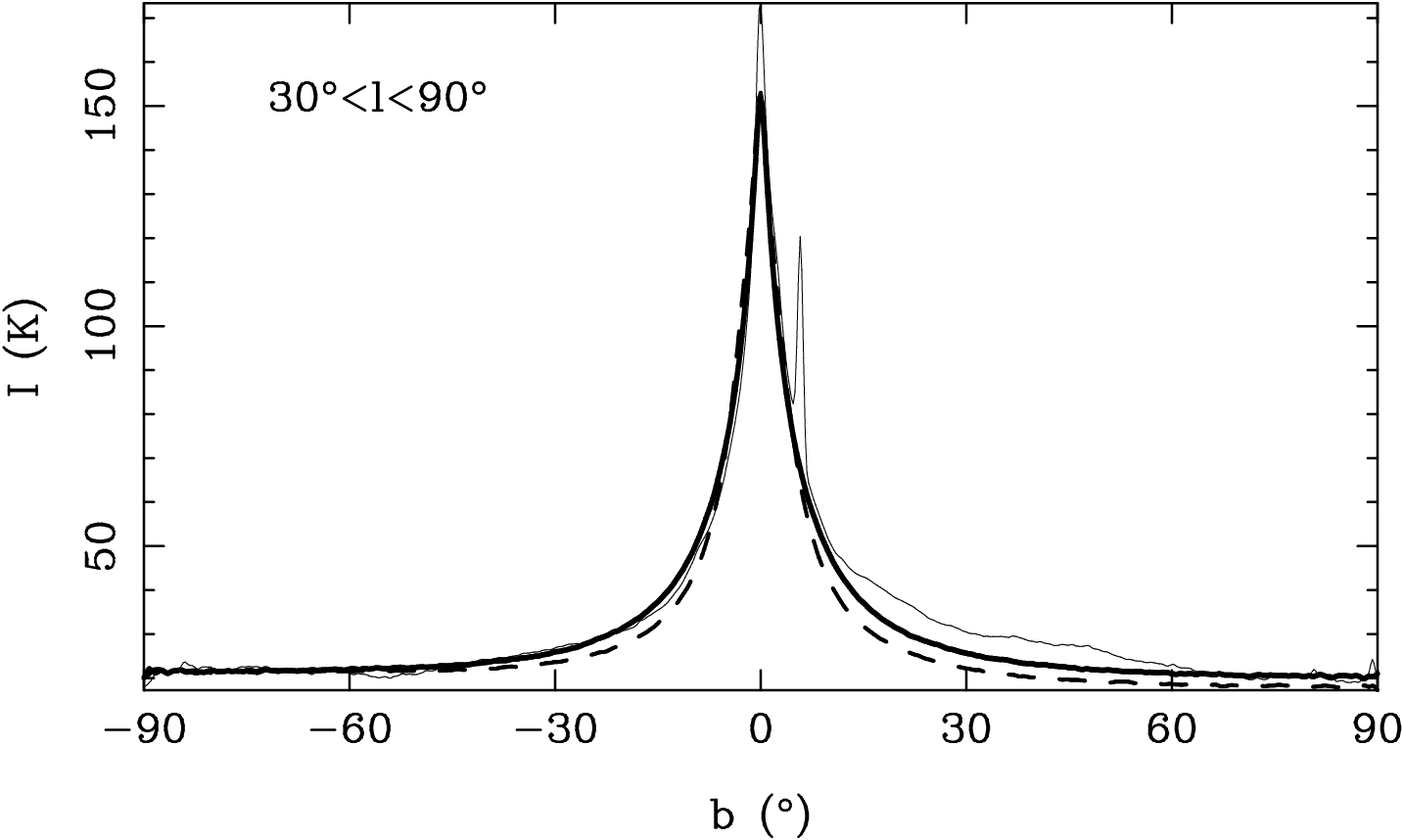}}
\resizebox{0.45\textwidth}{!}{\includegraphics[angle=-90]{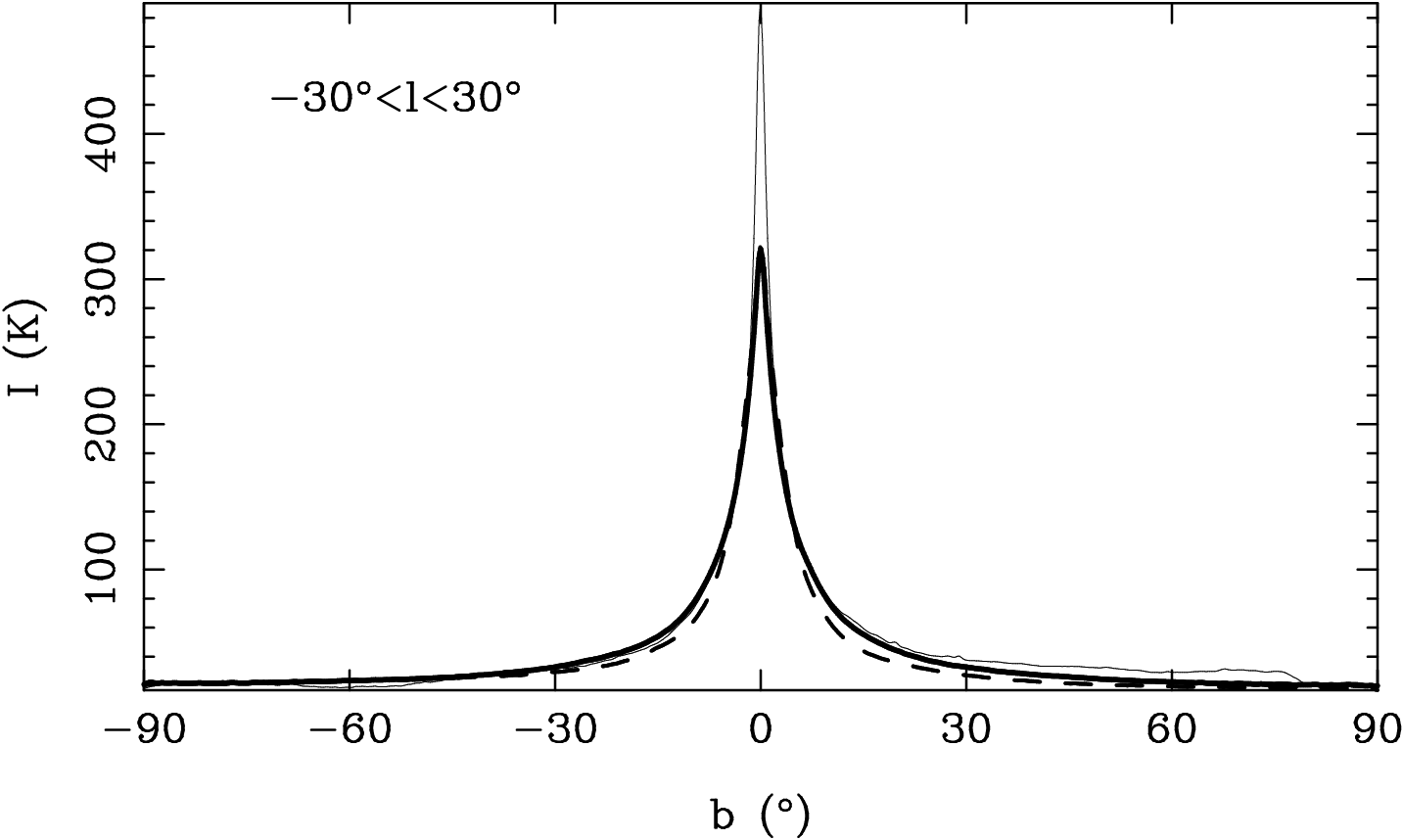}}
\resizebox{0.45\textwidth}{!}{\includegraphics[angle=-90]{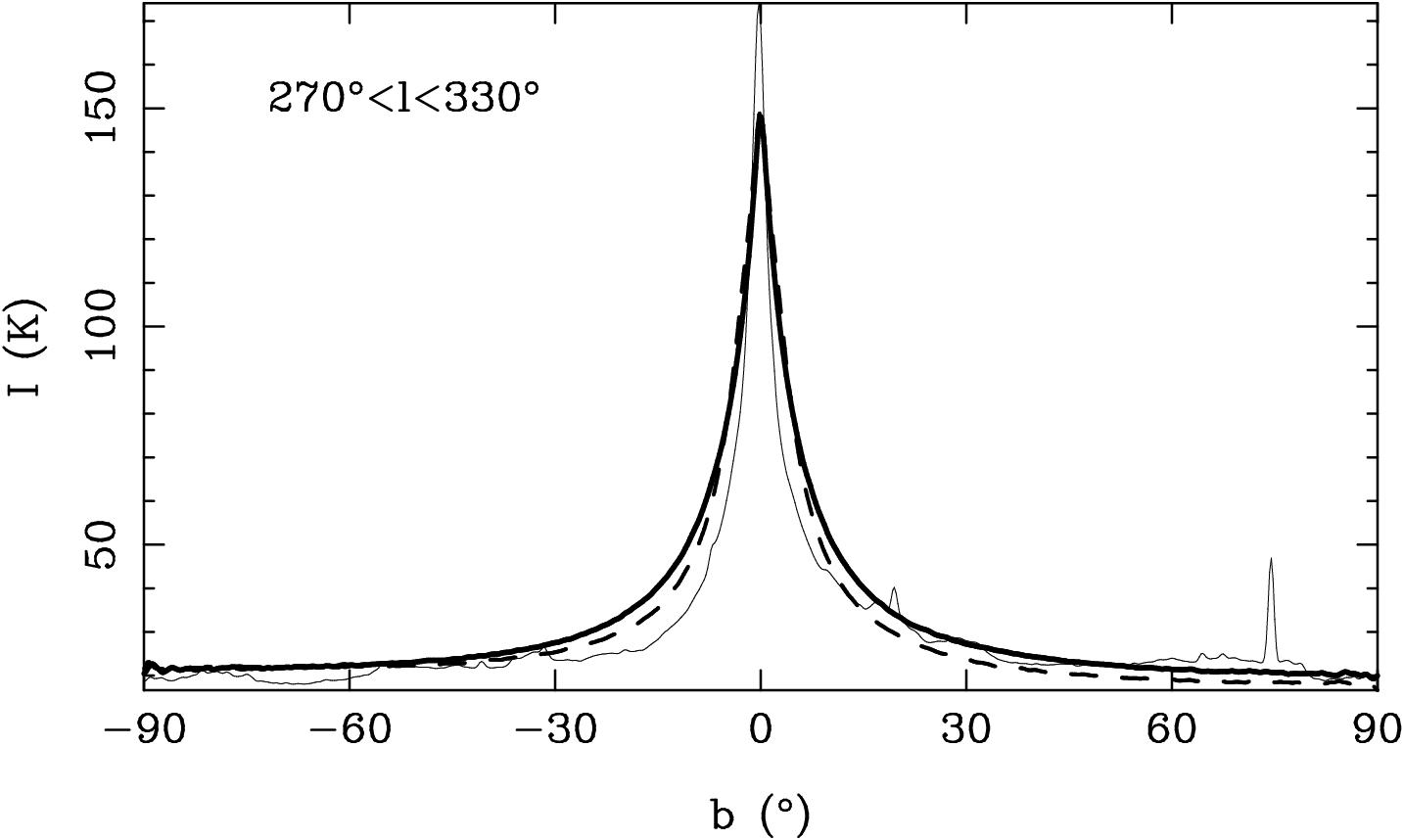}}
\resizebox{0.45\textwidth}{!}{\includegraphics[angle=-90]{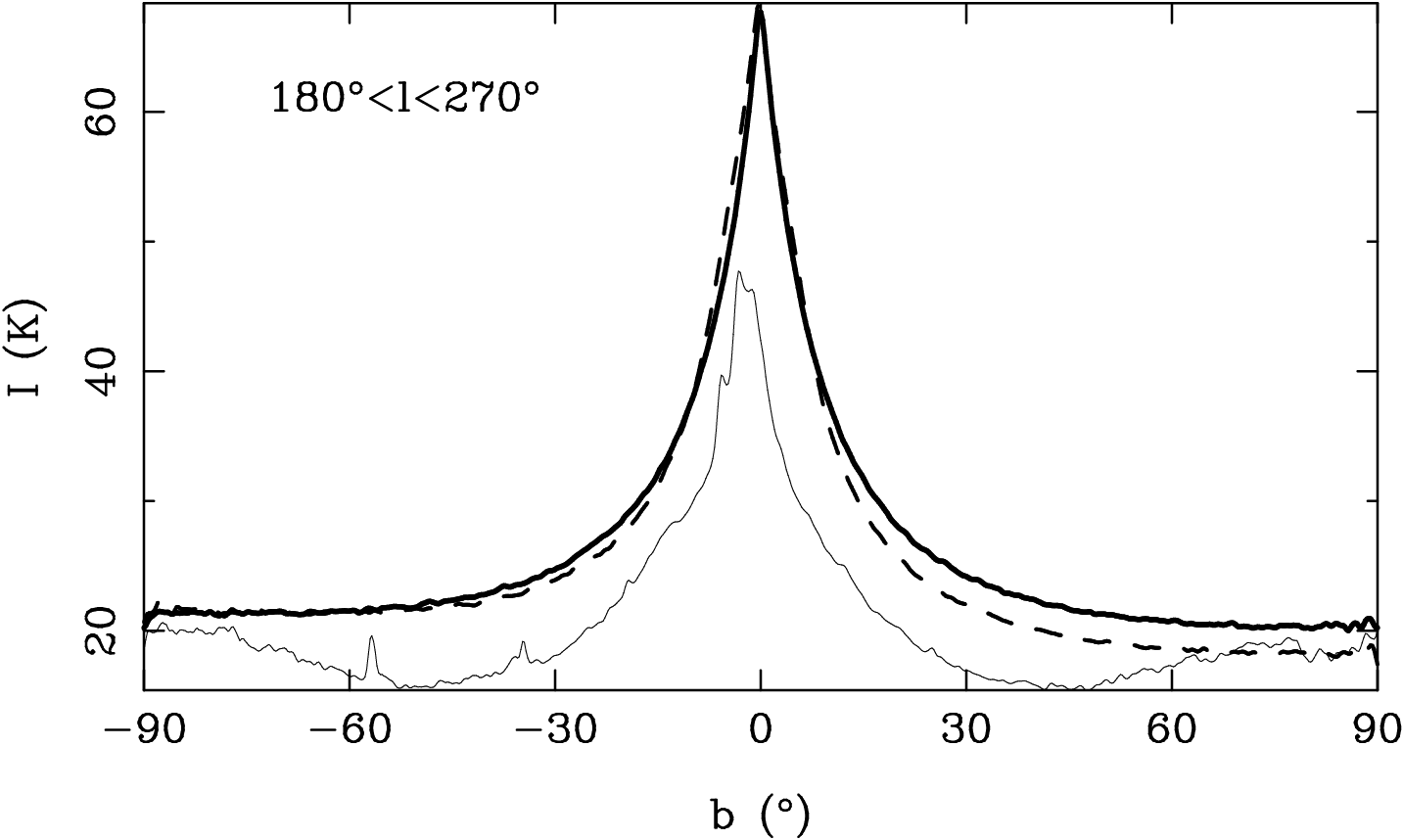}}
\caption{Total intensity profiles at 408~MHz. The thin solid lines show 
profiles obtained from the 408~MHz all-sky total intensity survey 
\citep{hssw82}. The results from the new model and the SRWE08 model are 
displayed as thick solid and dashed lines, respectively.}
\label{toti}
\end{figure}

\begin{figure}[!htbp]
\centering
\resizebox{0.45\textwidth}{!}{\includegraphics[angle=-90,width=6cm]{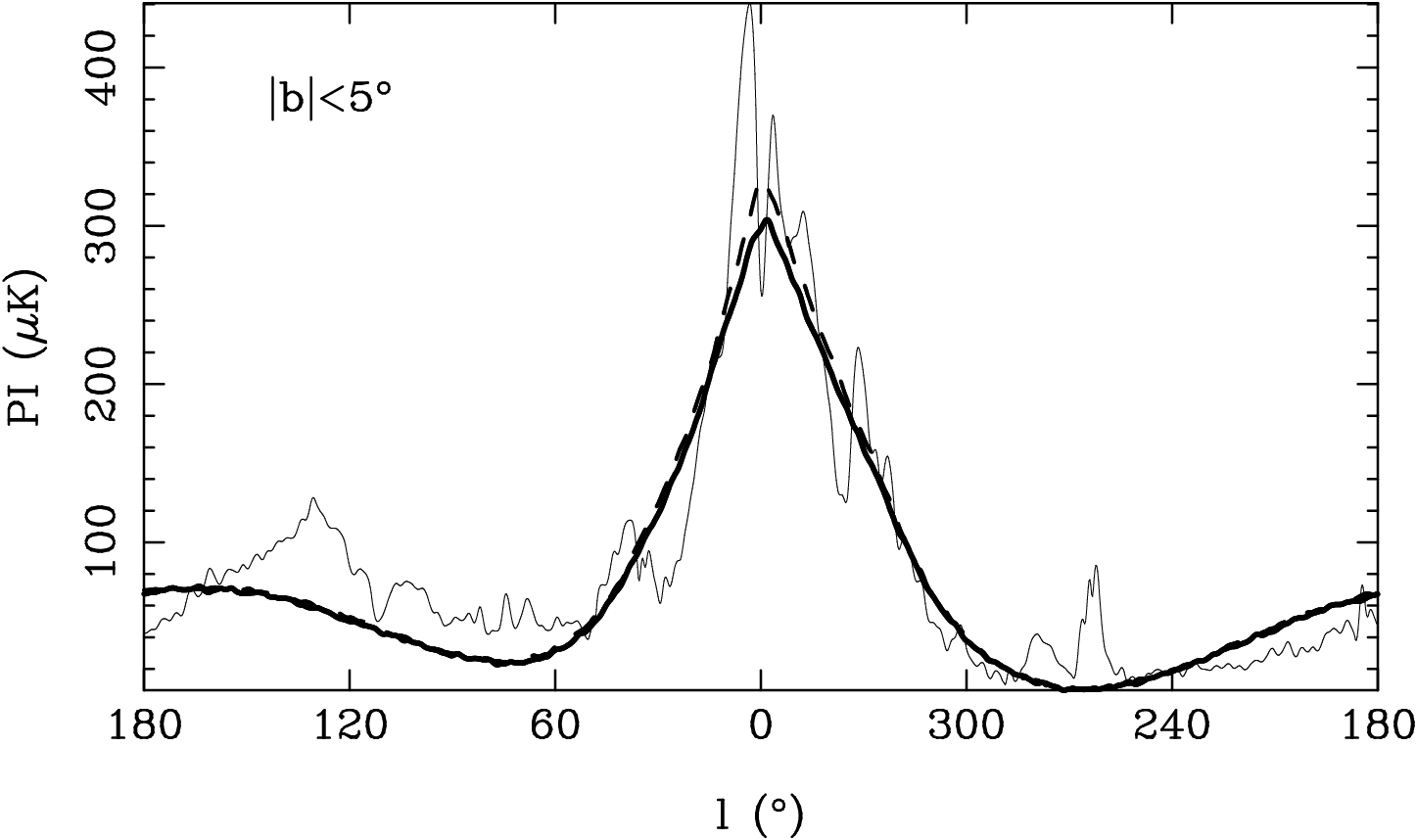}}
\resizebox{0.45\textwidth}{!}{\includegraphics[angle=-90,width=6cm]{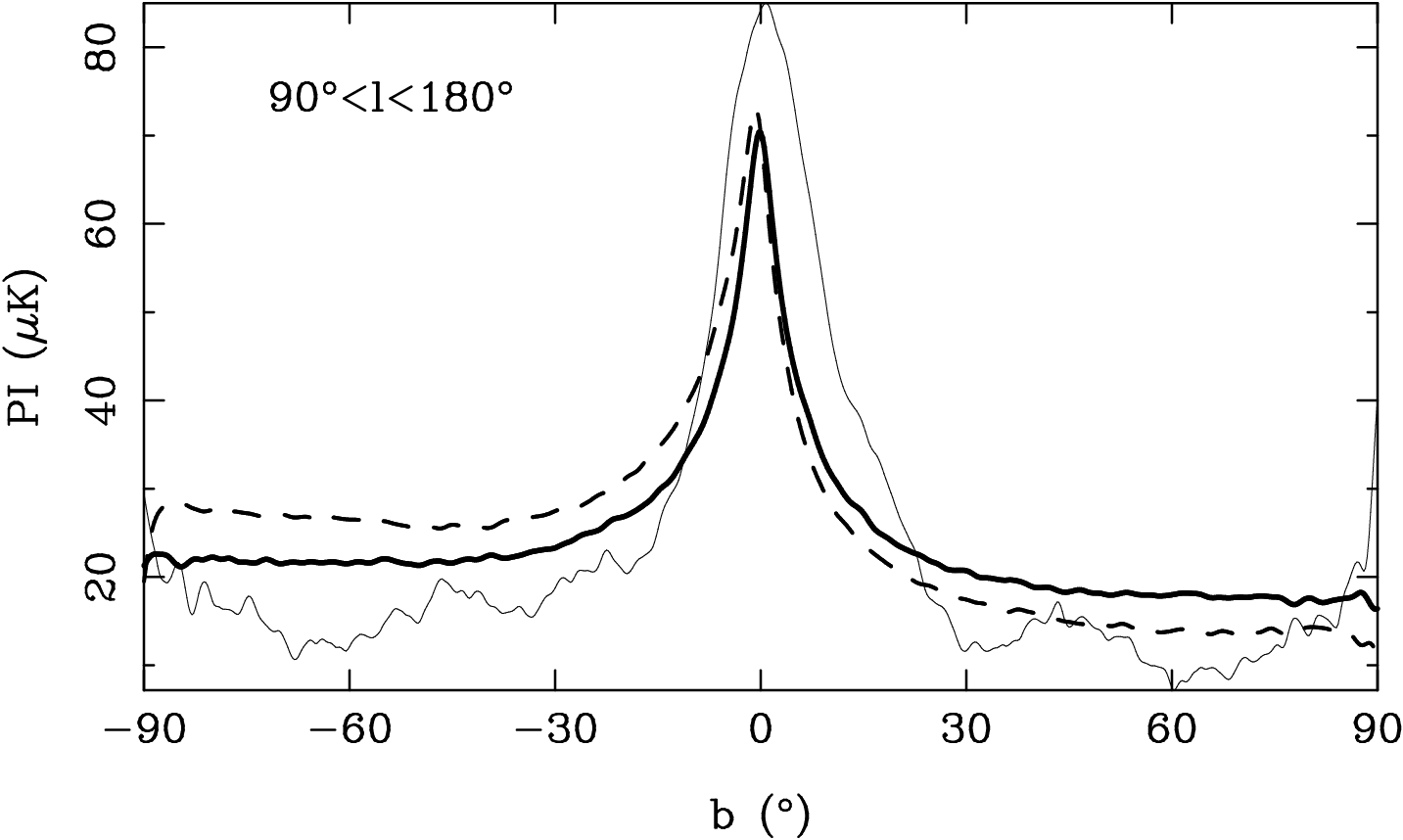}}
\resizebox{0.45\textwidth}{!}{\includegraphics[angle=-90,width=6cm]{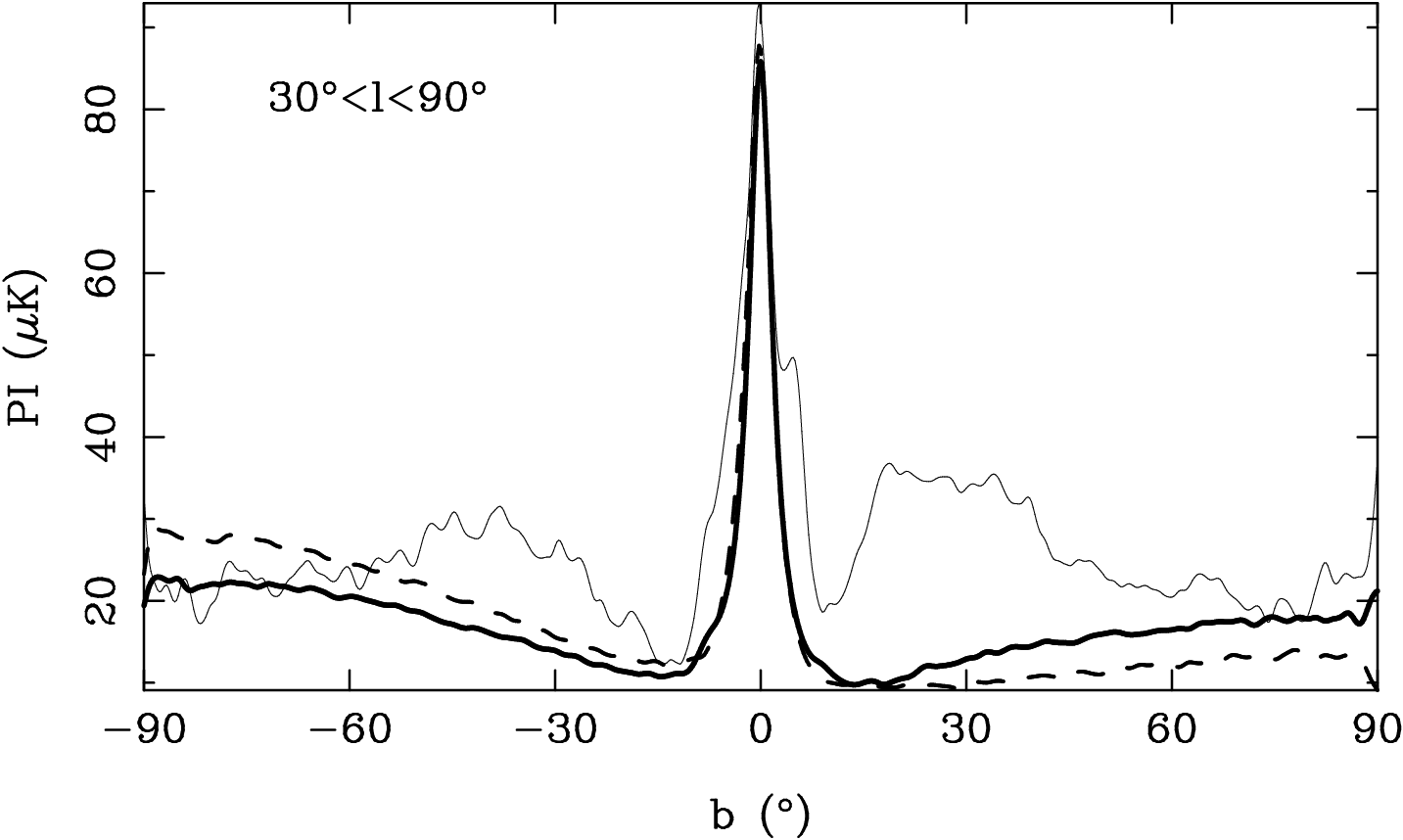}}
\resizebox{0.45\textwidth}{!}{\includegraphics[angle=-90,width=6cm]{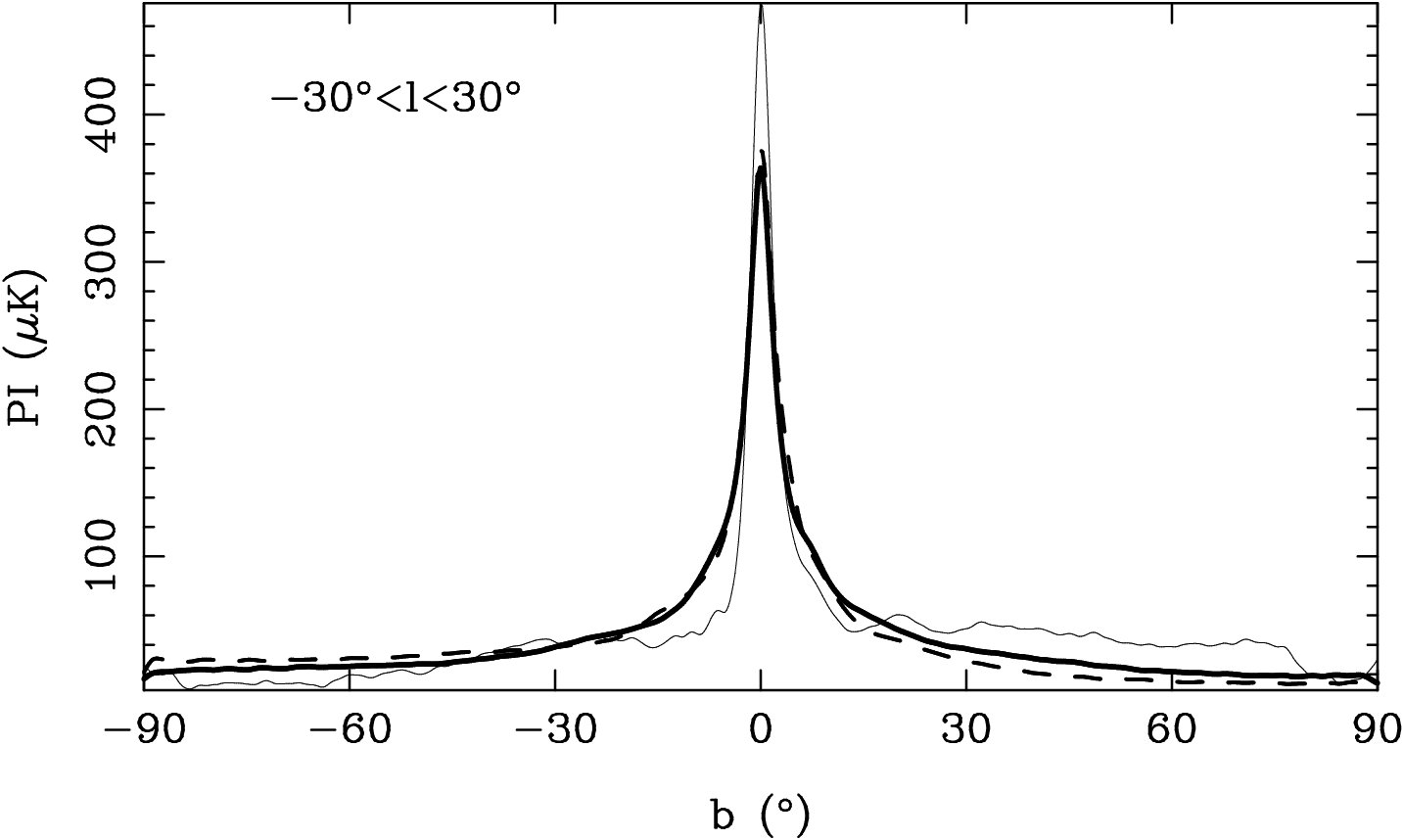}}
\resizebox{0.45\textwidth}{!}{\includegraphics[angle=-90,width=6cm]{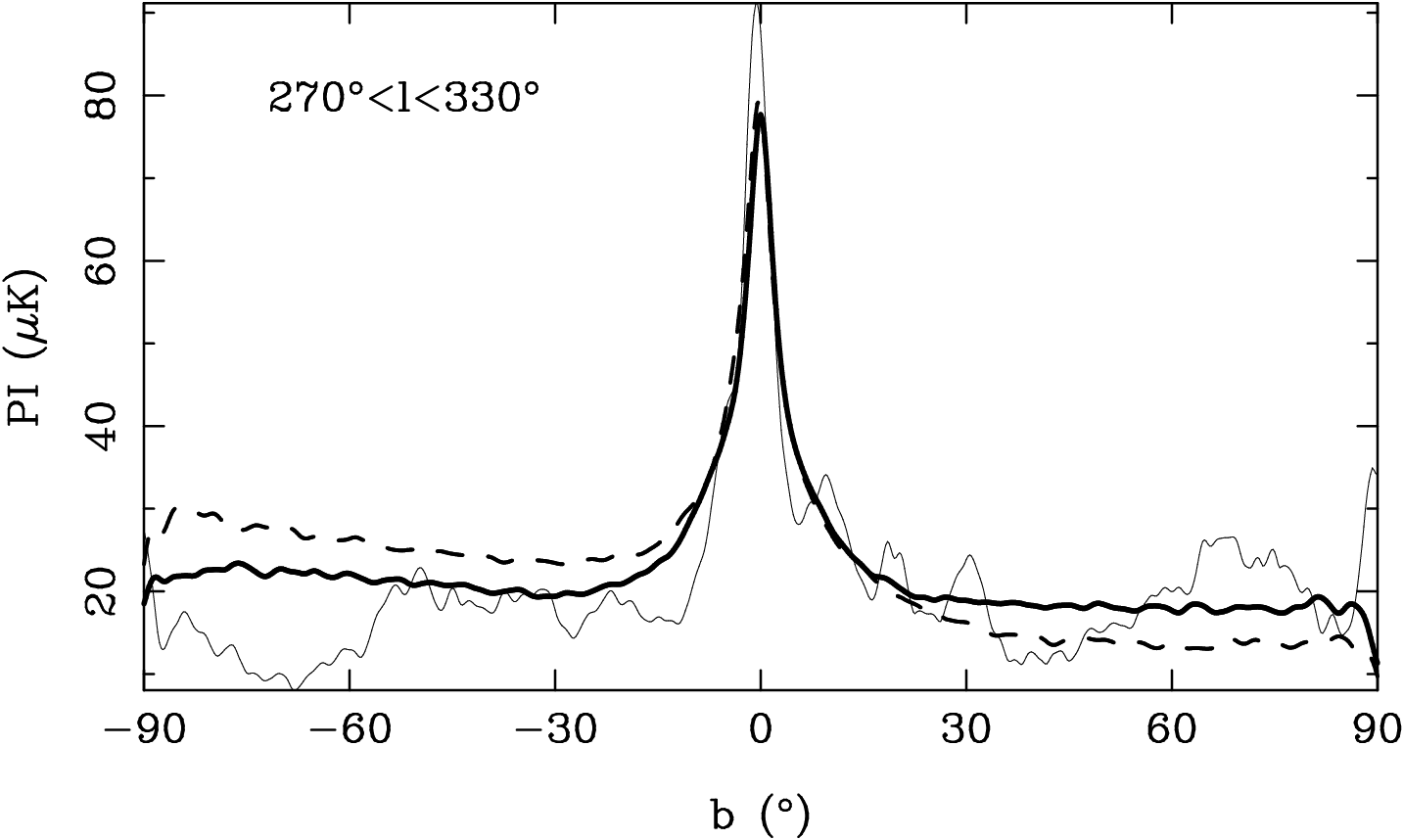}}
\resizebox{0.45\textwidth}{!}{\includegraphics[angle=-90,width=6cm]{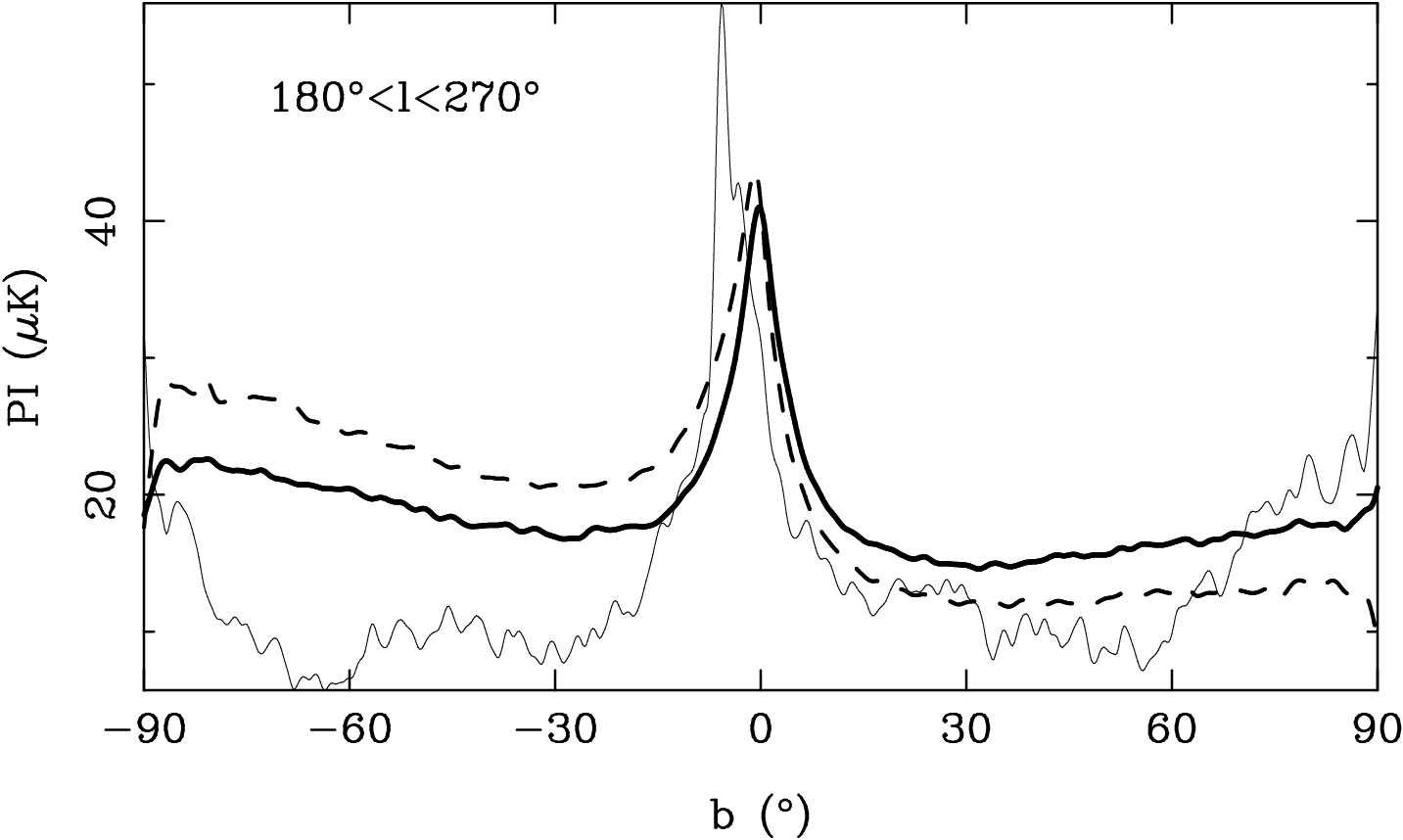}}
\caption{Profiles as in Fig.~\ref{toti}, but for 22.8~GHz polarized 
intensities from the WMAP five-years data \citep{hwh+09}.}
\label{ipiprof}
\end{figure}

\subsubsection{The cosmic-ray electron density distribution}

The SRWE08 models truncated the distribution of the cosmic-ray electron density 
at $|z|=1$~kpc to avoid excessive synchrotron emission at high latitudes caused 
by the strong halo magnetic field. This truncation is quite certainly 
physically unrealistic. Since the halo magnetic field in the revised models is 
much smaller this cutoff in $z$ is not needed any more. A scale-height of 
0.8~kpc instead of 1~kpc used by \citet{srwe08} was found to adapt better to 
the data.   

With the revised models, we simulated an all-sky total intensity map at 408~MHz 
and a polarized intensity map at 22.8~GHz at an angular resolution of 
$15\arcmin$. The 408~MHz map was then smoothed to $51\arcmin$, which is the 
angular resolution of the 408~MHz all-sky survey by \cite{hssw82}. The 
simulated as well as the WMAP 22.8~GHz polarization maps \citep{hwh+09} were 
smoothed to $2\degr$ to increase the signal-to-noise ratio. Slices extracted 
from the simulated maps are shown in Fig.~\ref{toti} for 408~MHz total 
intensities and in Fig.~\ref{ipiprof} for 22.8~GHz polarized intensities. The 
new simulations qualitatively agree with the observations at a level comparable 
to the SRWE08 models.

\subsubsection{The local enhancement}

\begin{figure}[!htbp]
\centering 
\resizebox{0.45\textwidth}{!}{\includegraphics[angle=-90]{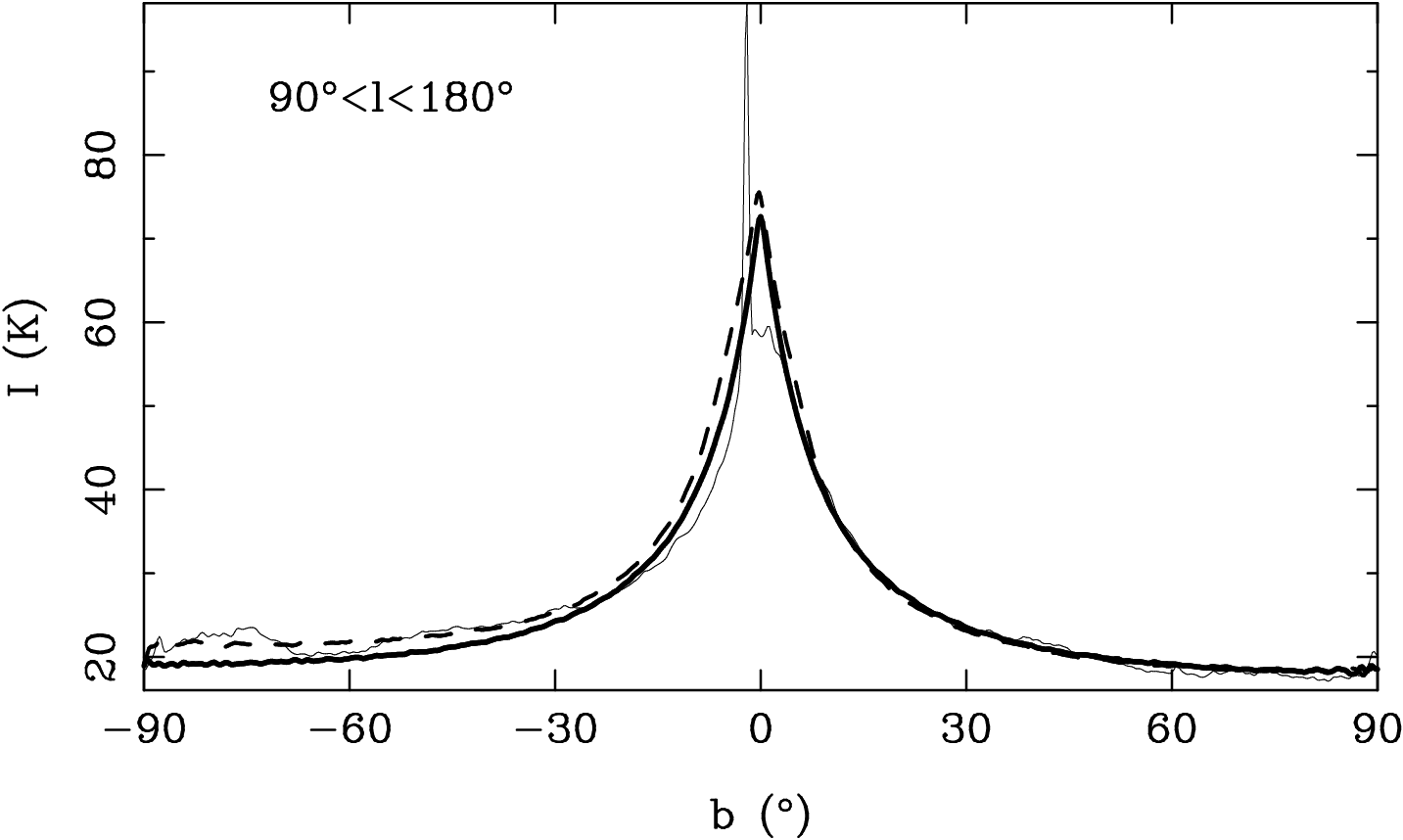}}
\resizebox{0.45\textwidth}{!}{\includegraphics[angle=-90]{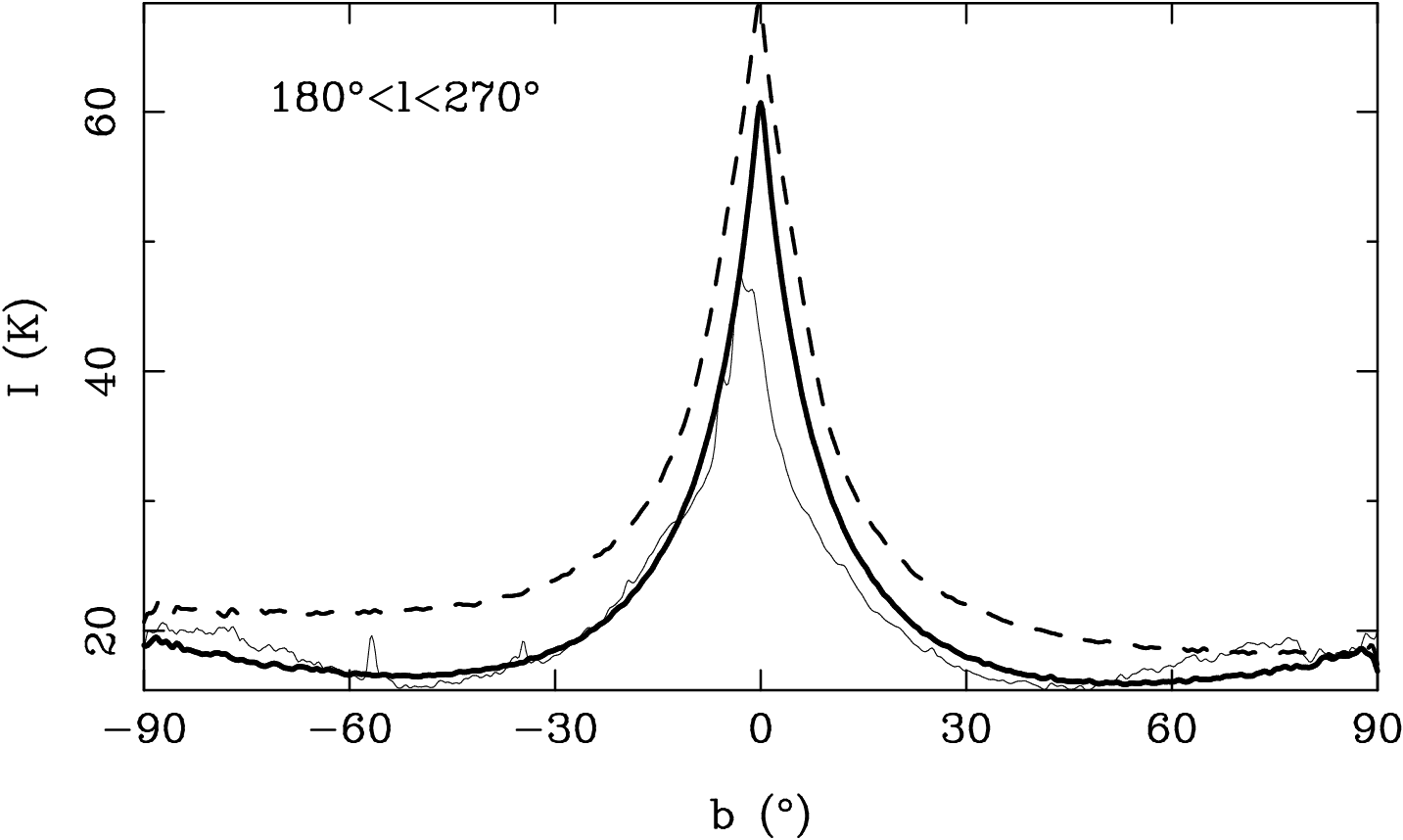}}
\caption{Total intensity latitude profiles. The thin lines are from the 408~MHz 
all-sky survey. The solid lines are from the simulations with the local 
enhancement shifting about 560~pc from the Sun, and the dashed lines are from 
the SRWE08 model.}
\label{totioff}
\end{figure}

The SRWE08 models include a local emission excess, realized by enhanced 
cosmic-ray electrons within 1~kpc, to account for the increase of the 
synchrotron emissivity towards the solar system based on low-frequency 
absorption data from optically thick \ion{H}{II} regions. Clearly the available 
data are quite limited and not suited to map the local synchrotron excess in 3D.
 The spherical approach used in the SRWE08 models is therefore almost 
arbitrary. It will be the task of LOFAR or other new low-frequency telescopes 
under construction to observe many more faint \ion{H}{II} regions across the 
sky to model the local synchrotron excess in 3D. This task is closely related 
to obtain a realistic halo model as both components are present at high 
latitudes. We note from Fig.~\ref{toti} that for the region of 
$180\degr<l<270\degr$ the observed 408~MHz total intensity at high latitudes is 
below that expected from the simulations. This indicates a possible offset of 
the local excess centre from the solar position. We shifted the centre of the 
local enhancement towards $l=45\degr$, $b=0\degr$ with a distance of about 
560~pc from the Sun and repeated our simulations. The results are shown in 
Fig.~\ref{totioff}. Quite obviously a much better agreement between the 
simulation and the observations is achieved shifting the centre of the local 
enhancement. However, there are not yet clues from observations on the centre 
location and the 3D geometry of the local enhancements, so that we are looking 
forward to numerous \ion{H}{II} region absorption results obtained by LOFAR.
 
\section{Summary}

In this paper we updated the parameters of the 3D-emissivity models used by 
\citet{srwe08} for all-sky simulations based on a revised thermal electron 
density scale-height by \citet{gmcm08}. The scale-height and mid-plane electron 
density of the NE2001 thermal electron density model of 1~kpc and 
0.034~cm$^{-3}$ were replaced by 1.8~kpc and 0.014~cm$^{-3}$. In consequence, 
the maximum halo field strength of 2~$\mu$G, and a scale-height of the 
cosmic-ray electron density of 0.8~kpc is obtained, which we consider to be 
much more physically relevant than the 10~$\mu$G field strength and the 
truncation with $z$ at 1~kpc obtained by \citet{srwe08}. With these 
modifications, we reproduce the high latitude RM distribution of 
extragalactic sources, total intensity and polarized intensity all-sky survey 
maps in the same way as \citet{srwe08}. We note that the mid-plane thermal 
electron density needs to be increased by a factor of about two to properly 
represent the RMs in the Galactic plane. We also find that a shift of the 
centre of the local synchrotron enhancement from the Sun position by 560~pc 
towards $l=45\degr$, $b=0\degr$ significantly improves the models to adapt to 
total intensity observations towards the outer Galaxy. 

\normalem
\begin{acknowledgements}
X.H. Sun acknowledges financial support by the MPG. He is grateful 
to Prof. M. Kramer for hospitality at the MPIfR and financial support.
We like to thank Patricia Reich and Rainer Beck for thorough reading of 
the manuscript and helpful discussions. 
\end{acknowledgements}

\label{lastpage}
\end{document}